\documentclass{aa}

\usepackage{graphicx}
\usepackage{txfonts}
\usepackage{orcidlink}

\newcommand{\tempotwo}{\textsc{tempo2}}
\newcommand{\psrchive}{\textsc{psrchive}}
\newcommand{\tnest}{\textsc{temponest}}

\begin{document}

\title{High-precision timing of seven binary millisecond pulsars with the European Pulsar Timing Array}

\author{K. Liu\orcidlink{0000-0002-2953-7376} \inst{1,2} \fnmsep\thanks{liukuo@shao.ac.cn}
          \and
          A. Chalumeau\orcidlink{0000-0003-2111-1001} \inst{3}
          \and
          H. Hu \orcidlink{0000-0002-3407-8071} \inst{4,2}
          \and
          S. Chen\orcidlink{0000-0002-3118-5963} \inst{1}
          \and
          L. Guillemot\orcidlink{0000-0002-9049-8716} \inst{5,6}
          \and
          J. Jang \inst{2}
          \and
          M. Keith \inst{7}
          \and
          J.~Antoniadis \inst{8,9}
          \and
          C.~Bassa \orcidlink{0000-0002-1429-9010}\inst{10}
          \and
          A.~Carleo \orcidlink{0000-0001-9929-2370}\inst{11}
          \and
          D.~J.~Champion \orcidlink{0000-0003-1361-7723}\inst{2}
          \and
          I.~Cognard\orcidlink{0000-0002-1775-9692} \inst{5,6}
          \and
          G.~H.~Janssen \orcidlink{0000-0003-3068-3677}\inst{10,12}
          \and
          W.~Jiang \inst{1}
          \and
          R. Karuppusamy \orcidlink{0000-0002-5307-2919} \inst{2}
          \and
          E. F. Keane \orcidlink{0000-0002-4553-655X} \inst{13}
          \and
          M.~Kramer \inst{2,7}
          \and
          Y.~Liu \inst{1}
          \and
          M.~B.~Mickaliger \orcidlink{0000-0001-6798-5682}\inst{7,14}
          \and
          A. Parthasarathy \orcidlink{0000-0002-4140-5616} \inst{10,15,2}
          \and
          D.~Perrodin \orcidlink{0000-0002-1806-2483}\inst{11}
          \and
          B. W. Stappers \inst{7}
          \and
          G.~Theureau\orcidlink{0000-0002-3649-276X} \inst{5,6}
          \and
          L.~Wang \orcidlink{0000-0003-0757-3584} \inst{1}
          }

   \institute{State Key Laboratory of Radio Astronomy and Technology, Shanghai Astronomical Observatory, CAS, 80 Nandan Road, Shanghai 200030, P. R. China
         \and
         Max-Planck-Institut f\"{u}r Radioastronomie, Auf dem H\"{u}gel 69, 53121, Bonn, Germany
         \and
         ASTRON, Netherlands Institute for Radio Astronomy, Oude Hoogeveensedijk 4, 7991 PD, Dwingeloo, The Netherlands
         \and
         Lohrmann Observatory, Technische Universit\"at Dresden, Mommsenstraße 13, 01062 Dresden, Germany
         \and
         LPC2E, OSUC, Univ Orleans, CNRS, CNES, Observatoire de Paris, F-45071 Orleans, France
         \and
         ORN, Observatoire de Paris, Universit\'e PSL, Univ Orleans, CNRS, 18330 Nan\c{c}ay, France Nan\c{c}ay, France
         \and
         Jodrell Bank Centre for Astrophysics, University of Manchester, Oxford Road, Manchester, M13 9PL, UK
         \and
         Institute of Astrophysics - Foundation for Research and Technology, Helas, N. Plastira 100, 70013, Heraklion, Greece
         \and
         Institute of Astrobiology, University of Crete
         \and
         ASTRON, Netherlands Institute for Radio Astronomy, Oude Hoogeveensedijk 4, Dwingeloo, 7991PD, The Netherlands
         \and
         INAF - Osservatorio Astronomico di Cagliari, Via della Scienza 5, 09047 Selargius, Italy
         \and
         Department of Astrophysics/IMAPP, Radboud University Nijmegen, P.O. Box 9010, 6500 GL Nijmegen, The Netherlands
         \and
         School of Physics, Trinity College Dublin, College Green, Dublin 2, D02 PN40, Ireland
         \and
         SKA Observatory, Jodrell Bank, Lower Withington, Macclesfield SK11 9FT, UK
         \and
         Anton Pannekoek Institute for Astronomy, University of Amsterdam, Science Park 904, 1098 XH Amsterdam, The Netherlands
}

   \date{Received; accepted}

  \abstract 
  {
  Binary millisecond pulsars (MSPs) serve as exceptional astrophysical laboratories for a wide range of astrophysical studies, all of which rely on precise property measurements allowed by pulsar timing. In this paper, we present high-precision timing solutions for seven binary MSPs monitored by the European Pulsar Timing Array. By combining data collected from the five largest radio telescopes in Europe as well as through their coherently combined form, the Large European Array for Pulsars, we achieved phase-coherent timing solutions for all seven pulsars. We conducted a Bayesian timing and noise analysis on the combined data, yielding measurements of astrometric and orbital parameters either for the first time or with broadly improved precision. For PSRs~J0125$-$2327, J1741+1351, and J2043+1711, we determined their distances and transverse velocities from astrometric measurements, and we inferred their masses via Shapiro delay measurements. Additionally, we examined the various contributions to the secular orbital-period variations for PSRs~J0125$-$2327, J1125+7819, and J1741+1351, and we estimated their kinematic distances. For PSRs J0125–2327 and J1741+1351, we detected the signature of annual-orbital parallax in the timing data. This allowed us to derive a strongly favoured, unique solution for the orbital ascending node and inclination angle for PSR J0125–2327 and a tightly constrained ascending node for PSR~J1741+1351. Moreover, we used the orbital periods and their derivatives for several binary MSPs to place upper limits on gravitational-wave strain in both the sub-nanohertz and microhertz frequency regimes. Our results show that the orbits of binary MSPs could provide stringent constraints in the microhertz range, thereby possibly complementing standard pulsar timing array analyses in the future.}

   \maketitle

\section{Introduction}
Millisecond pulsars (MSPs) are rapidly rotating neutron stars with spin periods of $\lesssim$ 30\,ms. They typically form through binary evolution \citep[e.g.][]{acrs82}. In this scenario the neutron star was initially born from core-collapse of the more massive star and subsequently accretes matter from its companion over gigayear timescales. This process transfers angular momentum and gradually spins up the neutron star, commonly resulting in a binary MSP with either a white dwarf or a neutron star companion \citep{tv23}. MSPs are renowned for their exceptional rotational stability, which enables their application as precise cosmic clocks \citep[e.g.][]{EPTA+2023a}. Via timing experiments with a baseline of years, binary MSPs can be utilised as sensitive probes of minuscule spacetime perturbations, making them outstanding tools for strong-field gravity experiments, such as tests of general relativity as well as constraints on effects predicted by alternative theories of gravity \citep[e.g.][]{afw+13,lgi+20,ksm+21,fw24}. Binary MSPs are also key components of pulsar timing array (PTA) experiments aimed at detecting nanohertz gravitational waves \citep[GWs; e.g.][]{EPTA+2023a,EPTA+2023c}. Moreover, their orbital dynamics can be exploited to search for GWs at other frequencies, such as in the picohertz and microhertz bands \citep{kop97,bj22a,bj22b}. Beyond studies of gravity, the observables of binary MSPs offer valuable insights into the evolutionary history of these systems and provide direct observational constraints on binary evolution models \citep[e.g.][]{imt+16,sfa+19,lgi+20}.

High-precision timing analysis allows for precise measurement of various astrometric and orbital parameters of binary pulsars \citep{lk04}. The astrometric parameters, including position, proper motion, and in some cases annual parallax, can be used to derive pulsar distances and transverse velocities. Pulsar distance in particular serves as a crucial input for a wide range of studies, such as testing gravitational theories, constraining the equation of state of ultra-dense matter using binary pulsars \citep[e.g.][]{hmn+20,vsb+25}, detecting continuous GWs from individual sources \citep[e.g.][]{bps+16,sbc+25}, and modelling the magnetic-field and free-electron distribution in the Milky Way \citep{xhj25,oc26}.

For binary orbits, timing solutions yield not only the five Keplerian orbital parameters but also, with sufficient data span and sensitivity, additional post-Keplerian (PK) parameters. These include secular variations in the Keplerian parameters as well as higher-order timing delays beyond the classical Roemer delay, such as the Shapiro and Einstein delays \citep{lk04}. When such PK measurements or their upper limits are dominated by relativistic orbital effects, they allow for determining the component masses of the binary and testing gravity. However, observed PK parameters often receive contributions from external influences, such as the Shklovskii effect, arising from the relative motion between the pulsar and the Solar System and the acceleration due to the Galactic gravitational potential \citep{shk70,McMillan:2017}. Therefore, a careful decomposition of all possible contributions to the measured PK parameters is essential before they can be robustly interpreted for gravity tests or mass measurements.

Here we present a detailed high-precision timing analysis of seven binary MSPs: PSRs~J0125–2327, J1125+7819, J1741+1351, J1911–1114, J2019+2425, J2033+1734, and J2043+1711. The data used in this study were collected by the European Pulsar Timing Array (EPTA), a European collaboration that operates five 100-m class radio telescopes dedicated to precision pulsar timing experiments \citep{EPTA+2023a}. The EPTA also operates the Large European Array for Pulsars (LEAP), which employs very long baseline interferometry techniques to coherently combine the five telescopes, achieving an equivalent diameter of up to 194 metres \citep{bjk+16}. The EPTA regularly monitors roughly 100 MSPs, a large fraction of which reside in binary systems, providing a rich dataset for studying individual binary MSPs. The seven pulsars selected for this work have been continuously monitored by the EPTA since as early as 2005. Three of them were included in the EPTA Data Release 1 \citep{dcl+16}, and all will be part of the upcoming EPTA Data Release 3.

The structure of this paper is as follows. In Section~\ref{sec:obs}, we describe the observations and data processing. Section~\ref{sec:results} presents the results of our precision timing analysis for all selected MSPs. In Section~\ref{sec:dis}, we discuss these timing results and use them to obtain additional properties of the pulsar systems. Finally, Section~\ref{sec:conclu} provides a brief summary and concluding remarks.

\section{Observations} \label{sec:obs}

The EPTA performs precision pulsar timing observations using the five largest European radio telescopes: the Effelsberg radio telescope (EFF) in Germany, the Lovell Telescope (LT) in the United Kingdom, the Nan\c{c}ay Radio Telescope (NRT) in France, the Sardinia Radio Telescope (SRT) in Italy, and the Westerbork Synthesis Radio Telescope (WSRT) in the Netherlands. These telescopes can also be operated in a phased-array mode known as LEAP \citep{bjk+16}. The first EPTA data release \citep[DR1;][]{dcl+16} comprised data from legacy recording systems, many of which employed incoherent dedispersion and had relatively narrow bandwidths of approximately 100 MHz. In contrast, the second data release \citep[DR2;][]{EPTA+2023a} incorporated data from a new generation of backends offering coherent dedispersion and substantially wider bandwidths of up to roughly 500 MHz. These newer systems rely on field-programmable gate array (FPGA) technology to digitise electrical signals and apply polyphase filterbanks. In most cases, data processing is carried out on the Reconfigurable Open Architecture Computing Hardware (ROACH) platform developed by the Collaboration for Astronomy Signal Processing and Electronics Research (CASPER; \citep{haa+16}). For the work presented in this paper, we utilised timing data from both the EPTA DR1 and the new-generation pulsar backends. In the following subsections, we briefly describe the data acquisition systems and the characteristics of the data products of the telescopes that directly contributed new-generation datasets to this study. Further details on the EPTA telescopes and instrumentation can be found in \citet{dcl+16} and \citet{EPTA+2023a}.

\begin{table*}
	\begin{center}
		\caption[]{Overview of the combined dataset and results from the timing analysis. \label{tab:data_overview}}
  \begin{tabular}{ccccccccccc}
  \multicolumn{8}{l}{\rule{18.9cm}{0.5pt}} \\
  Pulsar Jname & Telescopes & MJD range & $N_{\rm TOA}$ & Frequency & Median $\sigma_{\rm TOA}$ ($\mu$s) & wrms ($\mu$s) & wrms, whitened ($\mu$s) \\
\multicolumn{8}{l}{\rule{18.9cm}{0.5pt}} \\
J0125$-$2327 & EFF, NRT, LEAP &57928--60554 & 5352 & L,S & 0.409 & 0.367 & 0.294 \\
J1125+7819 & EFF, NRT & 56669--60537 & 4243 & L,S & 6.96 & 3.75 & -- \\
J1741+1351 & EFF, NRT & 55813--60551 & 875 & L,S & 2.75 & 0.905 & 0.805 \\
J1911$-$1114 & LT, NRT, DR1 & 53815--60240 & 422 & L & 9.63 & 4.35 & 3.92 \\
J2019+2425 & LT, NRT, DR1 & 53451--60505 & 511 & L & 22.8 & 10.5 & -- \\
J2033+1734 & LT, NRT, DR1 & 53898--60545 & 709 & L & 14.9 & 10.5 & -- \\
J2043+1711 & EFF, NRT & 55725--60533 & 310 & L & 1.53 & 0.629 & -- \\
\hline \noalign {\smallskip}
\label{tab:source}
\end{tabular}
\tablefoot{From the left to right, the columns give the pulsar name, the telescopes that have collected the data, the MJD range of the dataset, the total number of TOAs, the frequency band coverage (L: 1.0--2.0\,GHz; S: 2.0--4.0\,GHz), the median of TOA uncertainty ($\sigma_{\rm TOA}$), the weighted rms (wrms) of the timing residuals, and the wrms after subtracting the red noise components (if there is any).}
\end{center}
\end{table*}

\subsection{Effelsberg Radio Telescope}\label{ssec:eff}

The observations conducted by the Effelsberg 100-m radio telescope used in this paper were carried out in the L and S bands. The L-band data were acquired with the P200mm and P217mm receivers, centered at approximately 1.4\,GHz, providing bandwidths of 140 and 240\,MHz, respectively. The S-band observations were performed with the S110mm receiver at around 2.6\,GHz, with the observing bandwidth expanded from 80 to 300\,MHz in the second half of 2018. The typical integration time for each source was 30\,min, including a 2 min exposure on the noise diode for polarisation calibration. The ROACH-based `PSRIX' backend was employed to coherently de-disperse and fold the data streams in real time \citep{lkg+16}. Reduction of the folded archival data, including polarisation calibration, was carried out using the \textsc{CoastGuard} pipeline \citep{lkg+16}. Radio frequency interference (RFI) removal was performed with the \texttt{pazi} routine within the \psrchive{} software package \citep{vdo12}. Each epoch observation was averaged over time, frequency, and polarisation to produce integrated pulse profiles. The L-band data were divided into two sub-bands, except for PSR J2043+1711, for which the full band was used. A template profile was constructed for each pulsar and observing band by averaging the three observations with the highest signal-to-noise ratio (S/N). Times of arrival (TOAs) were then obtained by cross-correlating the pulse profiles with their corresponding templates using the traditional frequency-domain template-matching method \citep{tay92}.

\subsection{Lovell Telescope}\label{ssec:jbo}

The observations conducted by the 76-m Lovell Telescope typically ranged from 10 to 40\,min per source. A ROACH-based pulsar backend was used to sample data streams from two orthogonal polarisations at the Nyquist rate, digitising them as 8-bit integers. The full 512-MHz band was divided into 16-MHz sub-bands using a 32-channel polyphase filter. The signal in each sub-band was then coherently de-dispersed and folded at the pulse period in real time. Since November 2011, the data have additionally been cleaned for RFI using a spectral kurtosis method \citep{mrs22}. The lowest and highest 56\,MHz of the band were discarded due to known persistent RFI, leaving an effective bandwidth of 400\,MHz. Further RFI mitigation was performed on the archived data using a median filtering algorithm, followed by manual inspection. For each epoch observation, the archival data were fully averaged in time and partially averaged in frequency to produce two sub-bands centered at 1420 and 1620\,MHz, respectively. Template profiles were created by averaging a small number of high signal-to-noise ratio observations and removing white noise through multi-component fitting using the \texttt{paas} routine in \psrchive{}. The TOAs were generated for each sub-band using the \texttt{pat} routine in \psrchive{} and the \texttt{FDM} algorithm \citep{vlh+16}, with reference to the corresponding template profile at each frequency.

\subsection{Nan\c{c}ay Radio Telescope}\label{ssec:nrt}

The NRT conducts regular MSP timing observations using the L-band and S-band receivers, which provide frequency coverage of 1.1–-1.8 GHz and 1.7–-3.5 GHz, respectively. The typical observation duration per source ranges from 20 to 80\,min. Since August 2011, the Nan\c{c}ay Ultimate Pulsar Processing Instrument (NUPPI) has been the primary timing backend \citep{ctg+13,lgi+20}. Data were recorded over a 512-MHz bandwidth, channelised into 128 frequency channels, and coherently de-dispersed and folded in real time. The central frequency for L-band observations was 1484\,MHz, while for the S-band it was typically 2539\,MHz, although frequencies of 1854\,MHz or 2154\,MHz were occasionally used. Automatic RFI cleaning was performed on the archived data using a Python script based on the \texttt{surgical} RFI-cleaning algorithm from the \textsc{CoastGuard} software package \citep{lkg+16}. For each epoch, the pulse profiles were fully averaged in time and partially in frequency to produce four sub-bands while preserving the full set of four Stokes parameters. The TOAs were then measured from these polarimetric profiles using the matrix template matching technique, implemented via the \texttt{pat} routine in \psrchive{} \citep{van06}. Polarimetric templates were constructed for each individual frequency sub-band using the eight observations with the highest signal-to-noise ratios, following the procedures described in \citet{gcv+23}. The matrix template matching approach offers significantly improved characterisation of the receiver response and mitigates potential calibration artefacts, thereby yielding more accurate TOAs.

\subsection{The Large European Array for Pulsars}\label{ssec:leap}

Typically, LEAP coordinates monthly simultaneous observations at 1.4\,GHz using all five EPTA telescopes \citep{bjk+16}. In each session, the telescopes observe both the target pulsar and a nearby phase calibrator, recording baseband data at the Nyquist rate with 8-bit sampling across eight sub-bands of 16\,MHz each. The scan durations were typically of 45–60\,min for the pulsar and 2–3\,min for the calibrator. The data were then coherently combined using a dedicated software correlator to form a tied-array beam directed at the pulsar \citep{sbj+17}. The resulting baseband data were coherently de-dispersed and folded with the \textsc{dspsr} software package. Next, the archived data were visually inspected to remove RFI and subsequently averaged in time and frequency. For each pulsar, a template was constructed from the averaged profile of the observation with the highest signal-to-noise ratio, using the \texttt{psrsmooth} routine in \psrchive{} to suppress radiometer noise. The TOAs of each pulsar were extracted from the resulting profiles using the \texttt{pat} routine in \psrchive{} and the \texttt{FDM} algorithm \citep{vlh+16}.

\section{Results} \label{sec:results}

\begin{figure*}
\centering
\hspace*{-2cm}
	\includegraphics[scale=0.5]{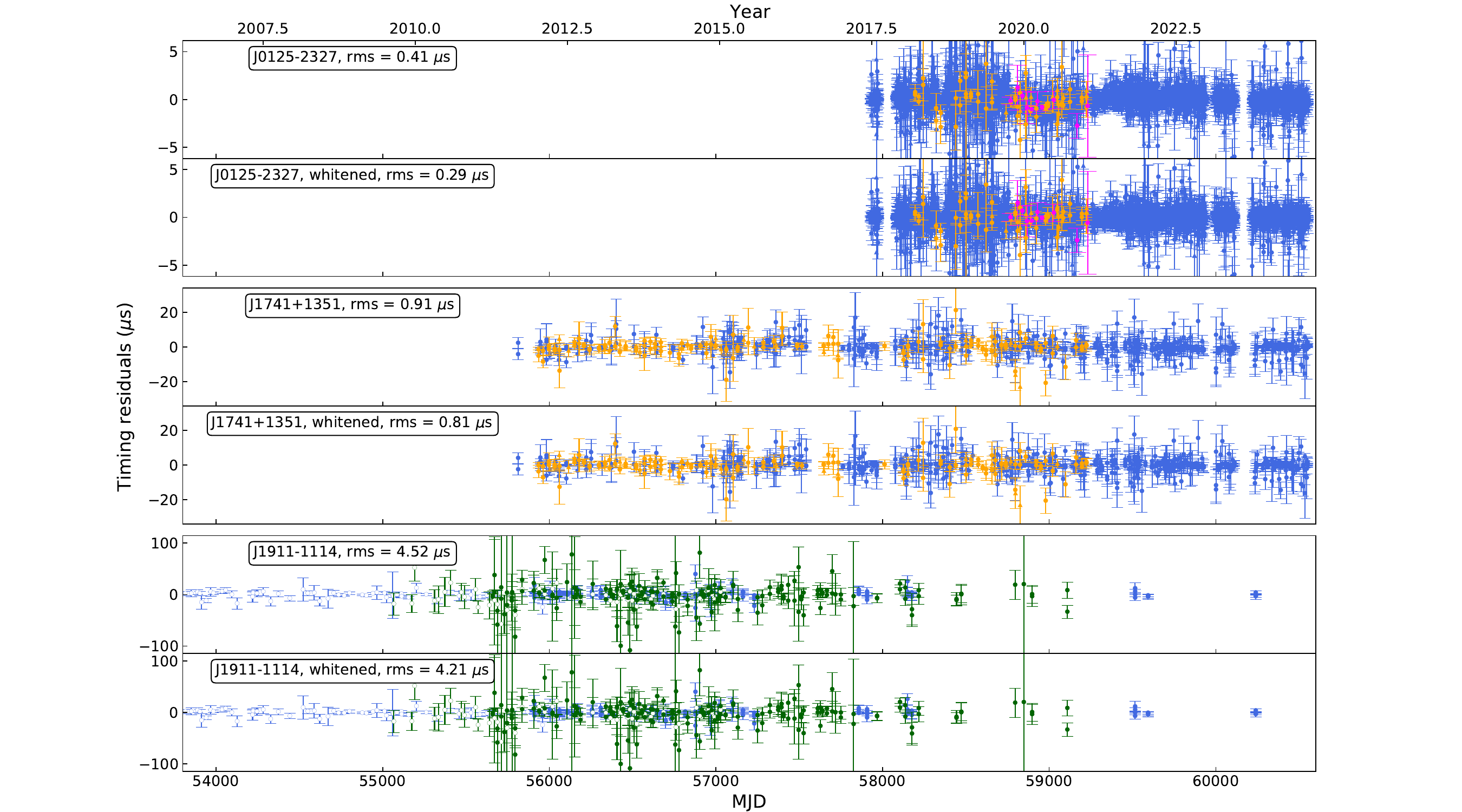}
    \caption{Post-fit timing residuals for PSRs~J0125$-$2327, J1741+1351, and J1911$-$1114, obtained from the Bayesian timing analysis in which the presence of red noise was detected in these pulsars. Timing residuals are shown both before and after subtraction of the red noise components. The data from EFF, LT, NRT, and LEAP are distinguished by orange, green, blue, and magenta colours, respectively. Circular and triangular symbols denote L-band and S-band observations, respectively. \label{fig:residual_rn}}
\end{figure*}

\begin{figure*}
\centering
\hspace*{-2cm}
	\includegraphics[scale=0.5]{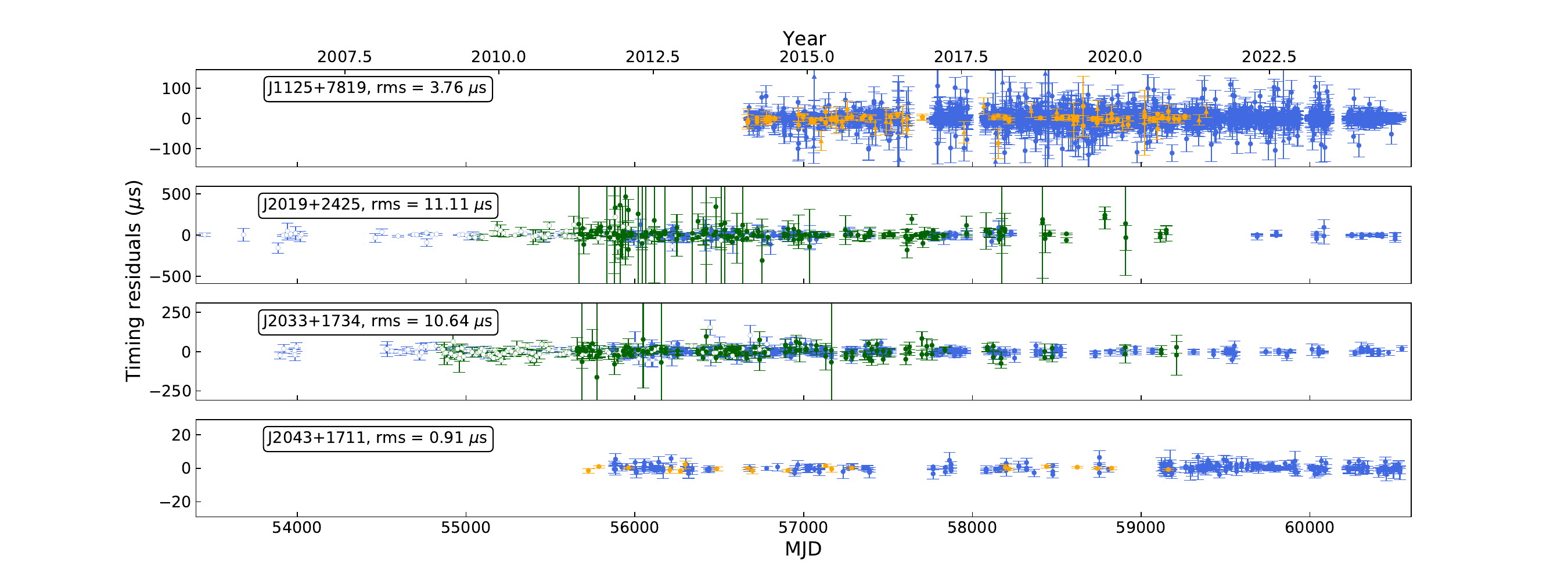}
    \caption{Post-fit timing residuals for PSRs~J1125+7819, J2019+2425, J2033+1734, and J2043+1711 obtained following the Bayesian timing analysis. The plotting conventions are the same as those in Figure~\ref{fig:residual_rn}.
    \label{fig:residual_norn}}
\end{figure*}

\begin{table*}[!htbp]
\centering
\caption{Measured timing and noise model parameters for PSRs~J0125$-$2327, J1125+7819, J1741+1351, and J1911$-$1114.}
\label{tab:ephem1}
\hspace*{-0.47cm}
\setlength{\tabcolsep}{4pt}
\begin{tabular}[c]{lllll}
\multicolumn{5}{l}{\rule{18.9cm}{0.5pt}} \\
Pulsar Jname & J0125$-$2327 & J1125+7819 & J1741+1351 & J1911$-$1114 \\
\multicolumn{5}{l}{\rule{18.9cm}{0.5pt}} \\
Reference epoch (MJD) & 58976 & 58602 & 58181 & 57027 \\
Right ascension, $\alpha$ (J2000) & 01:25:01.073826(1) & 11:25:59.88783(1) & 17:41:31.141393(2) & 19:11:49.27721(2) \\
Declination, $\delta$ (J2000) & -23:27:08.13684(2) & 78:19:48.70673(4) & 13:51:44.08170(5) & -11:14:22.534(2) \\
Spin frequency, $\nu$ (Hz) & 272.0810842741613(2) & 238.004049438789(1) & 266.8691624675484(9) & 275.805337857177(9) \\
Spin frequency derivative, $\dot{\nu}$ (s$^{-2}$) & -1.363396(4)$\times10^{-15}$ & -3.9333(2)$\times10^{-16}$ & -2.15209(3)$\times10^{-15}$ & -1.06528(10)$\times10^{-15}$ \\
DM (cm$^{-3}$~pc) & 9.5949(3)  & 11.216(3)  & 24.197(2) & 31.03(2) \\
DM1 (cm$^{-3}$~pc~yr$^{-1}$) & -0.00017(3) & ---  & -0.00037(4) & 0.0010(5) \\
DM2 (cm$^{-3}$~pc~yr$^{-2}$) & 3(4)$\times10^{-5}$ & --- & -5(2)$\times10^{-5}$ & -0.0005(2) \\
Proper motion in $\alpha$, $\mu_{\rm\alpha}$ (mas\,yr$^{-1}$) & 37.145(7) & 27.52(1) & -8.992(9) & -13.60(7) \\
Proper motion in $\delta$, $\mu_{\rm\delta}$ (mas\,yr$^{-1}$) & 10.639(10) & -2.18(1) & -7.50(1) & -9.8(4) \\
Parallax, $\varpi$ (mas) & 0.80(4) & --- & --- & --- \\
Binary model & ELL1H & ELL1 & ELL1H & ELL1  \\
Orbital period, $P_{\rm b}$ (d) & 7.27719975187(2)  & 15.355446197(1) & 16.3353480823(6) & 2.7165576625(3) \\
Projected semi-major axis, $x$ (s) & 4.729798963(6)  & 12.1924277(2) & 11.0033187(1) & 1.7628734(5) \\
Epoch of ascending node (MJD), $T_{\rm asc}$ & 59272.233608109(2) & 56157.47654600(9) & 57060.74133566(3) & 55888.7271085(1) \\
$\hat{x}$ component of the eccentricity, $\kappa$ & 6.5(3)$\times10^{-8}$ & -1.285(1)$\times10^{-5}$ & -3.89(1)$\times10^{-6}$  & 1.8(5)$\times10^{-6}$ \\
$\hat{y}$ component of the eccentricity, $\eta$ & 4.29(2)$\times10^{-7}$ & 9.0(1)$\times10^{-7}$ & -9.11(1)$\times10^{-6}$ & -4(6)$\times10^{-7}$ \\
Orbital period derivative, $\dot{P}_{\rm b}$ & 2.71(4)$\times10^{-12}$ & 2.6(4)$\times10^{-12}$ & 7(4)$\times10^{-13}$ & --- \\
Derivative of $x$, $\dot{x}$ & -3.85(1)$\times10^{-14}$ & -2.66(10)$\times10^{-14}$ & -2.7(6)$\times10^{-15}$ & --- \\
3rd harmonic of Shapiro delay, $h_{\rm 3}$ (s) & 6.4(4)$\times10^{-8}$ & --- & 5.4(4)$\times10^{-7}$ & --- \\
Ratio of harmonics amplitude, $\varsigma$ & 0.42(7) & --- & 0.70(5) & --- \\
Spectral amplitude of DM noise, $A_{\rm DM}$ & $-11.84_{-0.05}^{+0.05}$ & --- & --- & $-10.93_{-0.14}^{+0.12}$ \\
Spectral index of DM noise, $\gamma_{\rm DM}$ & $2.14_{-0.18}^{+0.19}$ & --- & --- & $0.48_{-0.32}^{+0.44}$ \\
Spectral amplitude of red noise, $A_{\rm red}$ & --- & --- & $-13.71_{-0.41}^{+0.34}$ & --- \\
Spectral index of red noise, $\gamma_{\rm red}$ & --- & --- & $3.38_{-0.92}^{+1.08}$ & --- \\
\multicolumn{5}{l}{\rule{18.9cm}{0.5pt}} \\
\end{tabular}
\tablefoot{The proper motion terms are defined as $\mu_\alpha = \dot{\alpha}\cos\delta$ and $\mu_\delta=\dot{\delta}$. The values shown in the table are the medians of the posterior distributions, and the brackets stand for the 1$\sigma$ credible interval of the respective parameters.}
\end{table*}

\begin{table*}[!htbp]
\centering
\caption{Continuation of Table~\ref{tab:ephem1} showing measurements of the timing and noise model parameters for PSRs~J2019+2425, J2033+1734, and J2043+1711.}
\label{tab:ephem2}
\hspace*{-0.47cm}
\begin{tabular}[c]{llll}
\multicolumn{4}{l}{\rule{15.9cm}{0.5pt}} \\
Pulsar Jname & J2019+2425 & J2033+1734 & J2043+1711 \\
\multicolumn{4}{l}{\rule{15.9cm}{0.5pt}} \\
Reference epoch (MJD) & 56978 & 57221 & 58129 \\
Right ascension, $\alpha$ (J2000) & 20:19:31.93718(3) & 20:33:27.51161(2) & 20:43:20.880021(3) \\
Declination, $\delta$ (J2000) & 24:25:14.9035(6) & 17:34:58.4678(4) & 17:11:28.86640(7) \\
Spin frequency, $\nu$ (Hz) & 254.16034137822(2) & 168.096675370623(7) & 420.189436597101(2) \\
Spin frequency derivative, $\dot{\nu}$ (s$^{-2}$) & -4.540(2)$\times10^{-16}$ & -3.1517(8)$\times10^{-16}$ & -9.2583(4)$\times10^{-16}$ \\
DM (cm$^{-3}$~pc) & 17.19(3) & 25.06(2) & 20.710(3) \\
DM1 (cm$^{-3}$~pc~yr$^{-1}$) & 0.003(1) & --- & --- \\
DM2 (cm$^{-3}$~pc~yr$^{-2}$) & --- & --- & --- \\
Proper motion in $\alpha$, $\mu_{\rm\alpha}$ (mas\,yr$^{-1}$) & -9.3(1) & -5.99(6) & -5.71(1) \\
Proper motion in $\delta$, $\mu_{\rm\delta}$ (mas\,yr$^{-1}$) & -20.3(2) & -9.5(1) & -10.84(2) \\
Parallax, $\varpi$ (mas) & --- & --- & 0.63(24) \\
Binary model & DD & DD & ELL1H \\
Orbital period, $P_{\rm b}$ (d) & 76.51163603(2) & 56.307796187(10) & 1.48229080943(1) \\
Projected semi-major axis, $x$ (s) & 38.767624(6) & 20.1631194(6) & 1.6239615(1) \\
Longitude of periastron, $\omega$ (deg) & 159.06(2) & 78.18(3) & --- \\
Epoch of periastron, $T_{\rm 0}$ & 50054.650(5) & 49878.139(4) & --- \\
Orbital eccentricity, $e$ & 0.00011103(5) & 0.00012864(6) & --- \\
Epoch of ascending node (MJD), $T_{\rm asc}$ & --- & --- & 55253.80385007(4) \\
$\hat{x}$ component of the eccentricity, $\kappa$ & --- & --- & -2.7(1)$\times10^{-6}$ \\
$\hat{y}$ component of the eccentricity, $\eta$ & --- & --- & -2.5(1)$\times10^{-6}$ \\
Derivative of $x$, $\dot{x}$ & 6(1)$\times10^{-14}$ & --- & --- \\
3rd harmonic of Shapiro delay, $h_{\rm 3}$ (s) & --- & --- & 5.7(7)$\times10^{-7}$ \\
Ratio of harmonics amplitude, $\varsigma$ & --- & --- & 0.77(7) \\
\hline
\end{tabular}
\end{table*}

For each pulsar, a timing analysis was carried out using a combination of data from different telescopes. To combine those data, we used the \tempotwo{} software package \citep{ehm06} and followed the procedures outlined in \citet{EPTA+2023a}. We adopted TT(BIPM2023) \citep{pet09}\footnote{\protect\url{https://webtai.bipm.org/ftp/pub/tai/ttbipm/}} as our reference clock standard and used the DE440 version of the JPL Solar System ephemeris \citep{pfwb21} to account for the Earth's orbital motion around the Sun. To align the data collected with different observing systems, we fitted for an arbitrary phase offset for each combination of receiver, backend and sub-band with respect to the NUPPI sub-band centred at 1420\,MHz. In cases where the data overlapped in time and frequency but were acquired with different systems, we retained the data from the newer backends. For each pulsar, we then performed a global fit for all known parameters to produce initial timing solutions. The inclusion of potential new timing parameters was initially assessed based on a 3$\sigma$ detection threshold and later verified through the results of the Bayesian timing analysis (see below). To correct for solar wind related DM variations, we employed the spherical solar wind electron density model. Our data did not allow for a constraint on the solar wind electron density at a 1-AU distance from the Sun. Since all pulsars in this study are located at an ecliptic latitude above 10\,deg, we fixed this value to 4\,cm$^{-3}$ in the model, following the recommendation of \citet{sct+24}.

To obtain the final timing solution for each pulsar, we performed a Bayesian timing analysis on the combined dataset using the \tnest{} software package \citep{lah+14} and the \textsc{multinest} nested sampler \citep{fhb+09}. For each pulsar, in addition to the deterministic timing-model parameters, a set of stochastic parameters were also included to characterise the noise present in the data. The white-noise component was parameterised by a multiplicative factor $E_{\rm f}$ (\texttt{EFAC}) accounting for possible miscalculation of TOA uncertainties and an additive factor $E_{\rm q}$ added in quadrature
(\texttt{EQUAD}) to model other noise contributions, such as pulse phase jitter and systematics \citep[e.g.][]{lvk+11,lkl+12}. The combined uncertainty is given by
\begin{equation}
    \sigma=\sqrt{E^2_{\rm q}+E^2_{\rm f}\sigma^2_{\rm r}},
    \label{eq:whitenoise}
\end{equation}
where $\sigma_{\rm r}$ is the measured TOA uncertainty. These pairs of noise parameters were introduced for each receiver and backend combination.

The red noise in the data was modelled using two components, accounting for chromatic and achromatic noise components, respectively. Both were treated as stationary stochastic processes with a power-law spectrum:
\begin{equation}
S(f)=\frac{A^2}{C_{\rm 0}}\left(\frac{f}{f_{\rm r}}\right)^{-\gamma},
\label{eq:rednoise}
\end{equation}
where $S(f)$, $A$, $\gamma$, and $f_{\rm r}$ correspond to the spectral
density of the power as a function of Fourier frequency, $f$; the spectral
amplitude; the spectral index; and the reference Fourier frequency (set to 1\,yr$^{-1}$), respectively. Here, the spectral amplitude is proportional to the observing radio frequency as $A\propto\nu^{-\alpha}$, where the chromatic index $\alpha$ is $2.0$ for the chromatic (mainly the DM variation) and achromatic noise components, respectively. Both spectra have a low-frequency cutoff determined by the inverse of the dataset time span. The constant $C_{\rm 0}$ was set to one for the chromatic component and $12\pi^2$ for the achromatic component. For each pulsar, we determined a customised noise model by following the procedures presented in \citet{EPTA+2023b}. This included using Bayesian model selection to decide on the red-noise components and the number of frequency bins used to sample the spectrum of each component. The timing and noise parameters were sampled simultaneously during the Bayesian timing analysis, while the phase offsets were analytically marginalised. Uniform priors were adopted for most model parameters, except for \texttt{EQUAD} and the red-noise amplitudes, for which log-uniform priors were used. The resulting measurements of timing and noise parameters are presented in Tables~\ref{tab:ephem1} and \ref{tab:ephem2}. Timing residuals obtained by applying the timing solutions are shown in Figure~\ref{fig:residual_rn} and Figure~\ref{fig:residual_norn}.

\subsection{PSR~J0125$-$2327}
PSR J0125-2327 has a spin period of 3.68\,ms and an orbital period of 7.28\,days \citep{mss+20}. Its orbital eccentricity is approximately $4.3\times10^{-7}$, making it the fourth smallest eccentricity among all known binary pulsars to date. The annual parallax was significantly detected in our timing analysis and is consistent with the measurement of \citet{sbf+24}, but the precision is improved by nearly a factor of three. In terms of orbital properties, the secular variation of the orbital period ($\dot{P}_{\rm b}$) and semi-major axis ($\dot{x}$) were measured with high significance. These quantities were later used to investigate the kinematic effects of the system (see Section~\ref{ssec:pbdot} and \ref{ssec:xdot}). Additionally, we obtained a significant measurement of the Shapiro delay characterised by the amplitude of the third harmonics ($h_{\rm 3}$) in the Fourier series and the amplitude ratio of the third and fourth harmonics ($\varsigma$) \citep{fw10}. These measurements were later used to constrain the masses of the two objects in the binary system (see Section~\ref{ssec:mass}). From the noise model selection analysis, we found the presence of DM noise in the dataset, while no preference was found for achromatic red noise. The achieved timing precision, after subtracting red noise, is just below 300\,ns, positioning this pulsar among the most precisely timed MSPs known.

\subsection{PSR J1125+7819}
PSR J1125+7819 has a spin period of 4.20\,ms and an orbital period of 15.4\,days \citep{lsk+18}. Both $\dot{P}_{\rm b}$ and $\dot{x}$ were measured in our timing analysis. Notably, this is the first reported measurement of $\dot{P}_{\rm b}$ for this pulsar. Our measurement of $\dot{x}$ is consistent with that reported by \citet{NG15+a}, but the precision is improved by a factor of four.

\subsection{PSR J1741+1351}
PSR J1741+1351 has a spin period of 3.75\,ms and an orbital period of 16.3\,days \citep{jbo+07}. For this pulsar, $\dot{P}_{\rm b}$ and $\dot{x}$ were only marginally constrained, while the Shapiro-delay effect was measured with high significance. These measurements are in general consistent with those reported by \citet{NG15+a}. The noise model selection analysis revealed the presence of achromatic red noise in the data with no preference for DM noise.

\subsection{PSR J1911$-$1114}
PSR J1911$-$1114 has a spin period of 3.63\,ms and an orbital period of 2.72\,days \citep{llb+96}. Our analysis yielded a marginal constraint on secular variation for one of the orthogonal eccentricity components ($\hat{y}$). However, we excluded this parameter from the final timing solution since its value was likely to be unphysical. The noise model selection analysis indicates a preference for DM noise in the data, with no evidence of achromatic red noise.

\subsection{PSR J2019+2425}
PSR J2019+2425 has a spin period of 3.93\,ms and an orbital period of 76.5\,days \citep{ntf93}. This pulsar has served as an important test bed for binary evolution studies, owing to its large magnetic inclination angle and old age \citep{tau98}. Our timing analysis yielded a measurement of $\dot{x}$, which is consistent with the value reported in \citet{nss01}.

\subsection{PSR J2033+1734}
PSR J2033+1734 has a spin period of 5.95\,ms and an orbital period of 56.3\,days \citep{rtj+96}. Assuming a 1.35-M$_{\odot}$ pulsar mass, the mass function implies the presence of a white dwarf companion with a mass of 0.19--0.22\,M$_{\odot}$ for an orbital inclination of 60--90\,deg. However, binary evolution models predict a white dwarf mass of 0.30--0.33\,M$_{\odot}$ for the given orbital period \citep{tv23}. This discrepancy suggests that the system may have a low orbital inclination or an unusually massive neutron star. For instance, adopting a white dwarf mass of 0.315\,M$_{\odot}$ would correspond to an orbital inclination angle of approximately 39\,deg or, alternatively, a pulsar mass of roughly 2.4\,M$_{\odot}$. Our timing analysis did not detect a clear Shapiro delay signal, which is consistent with the findings of \citet{NG15+a} and supports the scenario of a low-inclination orbit.

\subsection{PSR J2043+1711}
PSR J2043+1711 has a spin period of 2.37\,ms and an orbital period of 1.48\,days. It was discovered in a pulsar search from Fermi Large Area Telescope sources by the NRT \citep{gfc+12}. Our timing analysis yielded a marginal parallax measurement that is consistent with the value reported by \citet{NG15+a}. The Shapiro delay effect was significantly detected, again characterised by $h_{\rm 3}$ and $\varsigma$. These measurements translated into Shapiro delay measurements of $\sin{i}$ and $m_{\rm c}$ that are consistent with those reported in \citet{NG15+a}.

\section{Discussions} \label{sec:dis}
\subsection{Annual parallax and pulsar distance} \label{ssec:parallax}

\begin{table*}
\begin{center}
\caption{Parallax, distance, and velocity measurements of PSRs~J0125$-$2327, J1741+1351, and J2043+1711.}
\renewcommand{\arraystretch}{1.4}
\centering
\begin{tabular}{cccccccccc}
\hline
Pulsar name & $\varpi_{\rm mes}$ (mas) & F (mJy) & $\varpi_{\rm corr}$ (mas) & $D_{\rm corr}$ (kpc) & $D_{\rm Y}$, $D_{\rm N}$ (kpc) & $D_{\rm kin}$ (Kpc) & $\mu$ (mas\,yr$^{-1}$) & $V_{\rm trans}$ (km\,s$^{-1}$) \\
\hline
J0125$-$2327 & $0.80^{+0.04}_{-0.04}$ & $2.49^{+0.12}_{-0.12}$  & $0.79^{+0.04}_{-0.04}$ & $1.24^{+0.06}_{-0.06}$ & 0.876, 0.643 & $1.24^{+0.02}_{-0.02}$ & $38.638^{+0.007}_{-0.007}$ & $227^{+11}_{-11}$ \\
J1741+1351 & $0.41^{+0.10}_{-0.10}$ & $0.29^{+0.04}_{-0.04}$ & $0.38^{+0.09}_{-0.08}$ & $2.07_{-0.33}^{+0.45}$ & 1.36, 1.02 & $1.79_{-0.78}^{+0.81}$ & $11.710^{+0.009}_{-0.009}$ & $115^{+25}_{-18}$ \\
J2043+1711 & $0.64^{+0.23}_{-0.23}$ & $0.121^{+0.014}_{-0.014}$ & $0.40^{+0.17}_{-0.12}$ & $1.37^{+0.49}_{-0.31}$ &1.47, 1.77 & ---  & $12.25^{+0.02}_{-0.02}$ & $80^{+28}_{-18}$ \\
\hline
\end{tabular}
\tablefoot{From left to the right, the columns show the pulsar name, annual parallax measurement, flux density at the L-band frequency, parallax and distance after correcting for the LK bias, distances derived from the \textsc{YMW16} and \textsc{NE2025} Galactic free-electron density model, kinematic distance, proper motion, and transverse velocity of the pulsar system. Here, the parallax value of PSR J1741+1351 was obtained from \citet{NG15+a}. The flux densities of the pulsars were obtained from \citet{sbm+22}. The kinematic distances were directly taken from Table~\ref{table:pbdot} for comparison purposes.}
\label{table:parallax}
\end{center}
\end{table*}

We used the annual parallax measurements obtained from our timing analysis for PSRs J0125–2327 and J2043+1711 to derive their distances from Earth. To obtain accurate distance estimates, we applied a correction for the Lutz–Kelker (LK) bias, which arises from the asymmetric volume of space sampled on either side of the measured parallax value \citep{vlm10,vwc+12}.\footnote{The software routine used to account for the LK bias is available at \url{http://psrpop.phys.wvu.edu/LKbias/}.} The volumetric and pulsar luminosity priors adopted in this correction were the same as those used in \citet{vwc+12}, ensuring that the posterior distance distribution yields well-defined statistical moments even in the presence of large uncertainties. For the distance from Earth to the Galactic centre, we adopted a value of 8.28\,kpc, as reported in \citet{GRAVITY:2021}. Combining the parallax-derived distances with the proper motion measurements, we computed the transverse velocities of these pulsars. The same analysis was also applied to PSR J1741+1351 using the parallax measurement reported in \citet{NG15+a}.

A summary of the results is presented in Table~\ref{table:parallax}, which also includes distance estimates derived from the \textsc{YMW16} and \textsc{NE2025} Galactic free-electron models for comparison \citep{ymw17,oc26}. For PSRs J1741+1351 and J2043+1711, the distances are generally consistent across measurements. In contrast, for PSR J0125–2327, the values differ significantly. Given that this pulsar is located at a Galactic latitude of approximately -81.6\,deg, this discrepancy suggests that the Galactic free-electron density models still require further improvement in their distance predictions at very high Galactic latitudes.

\subsection{Mass measurements} \label{ssec:mass}

\begin{figure}
\hspace{-0.5cm}
\vspace{0.3cm}
\includegraphics[scale=0.45]{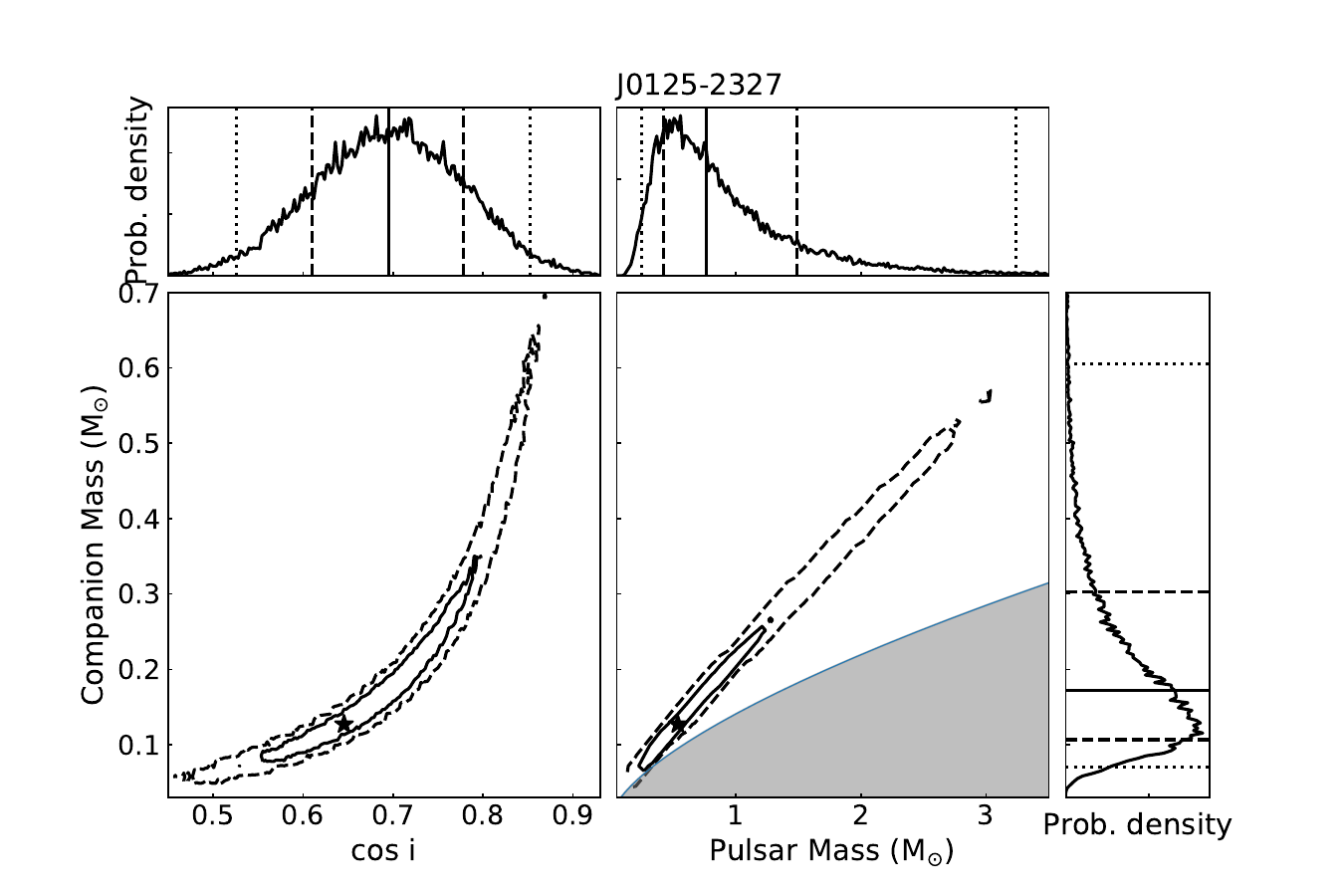}

\hspace{-0.5cm}
\vspace{0.3cm}
\includegraphics[scale=0.45]{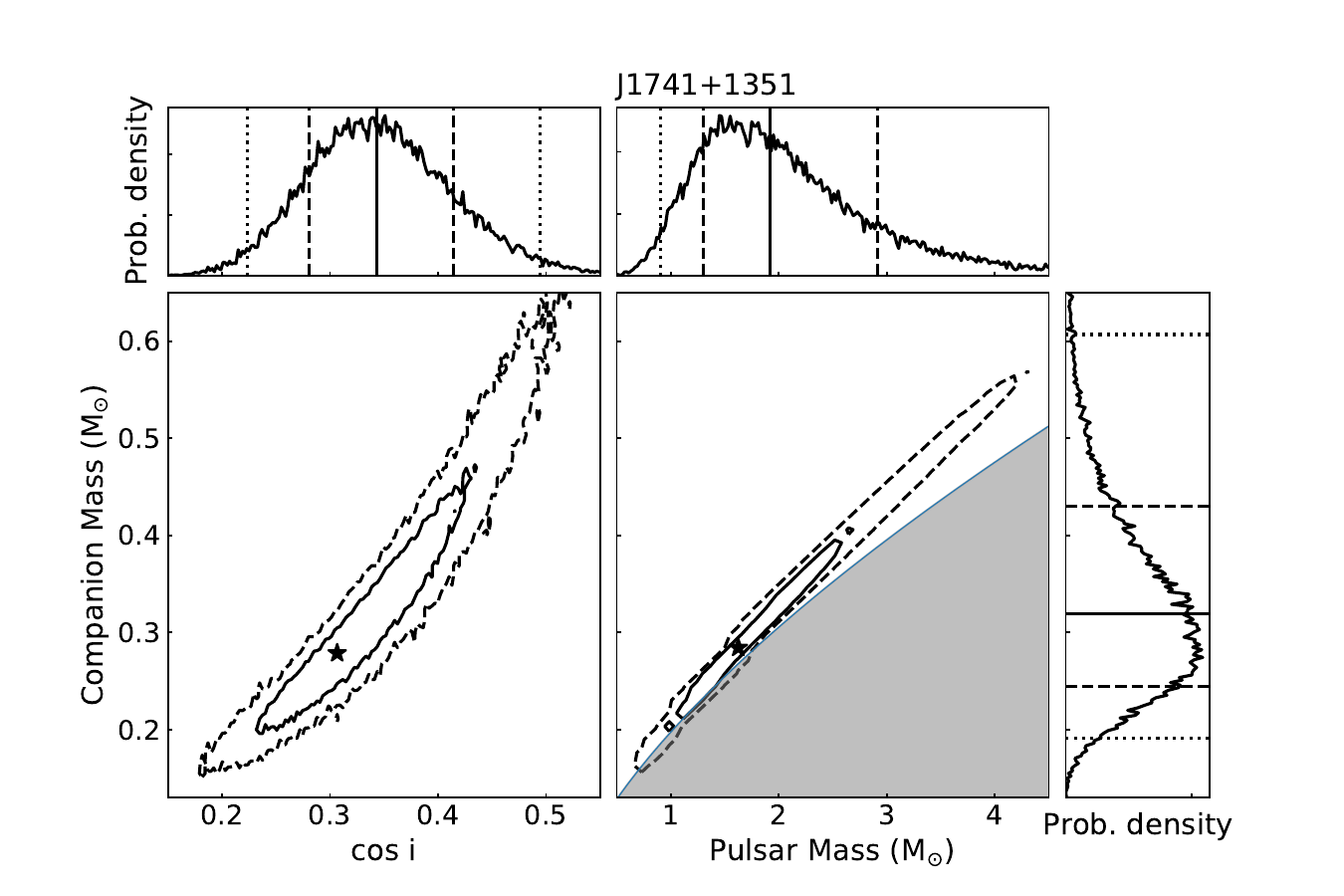}

\hspace{-0.5cm}
\includegraphics[scale=0.45]{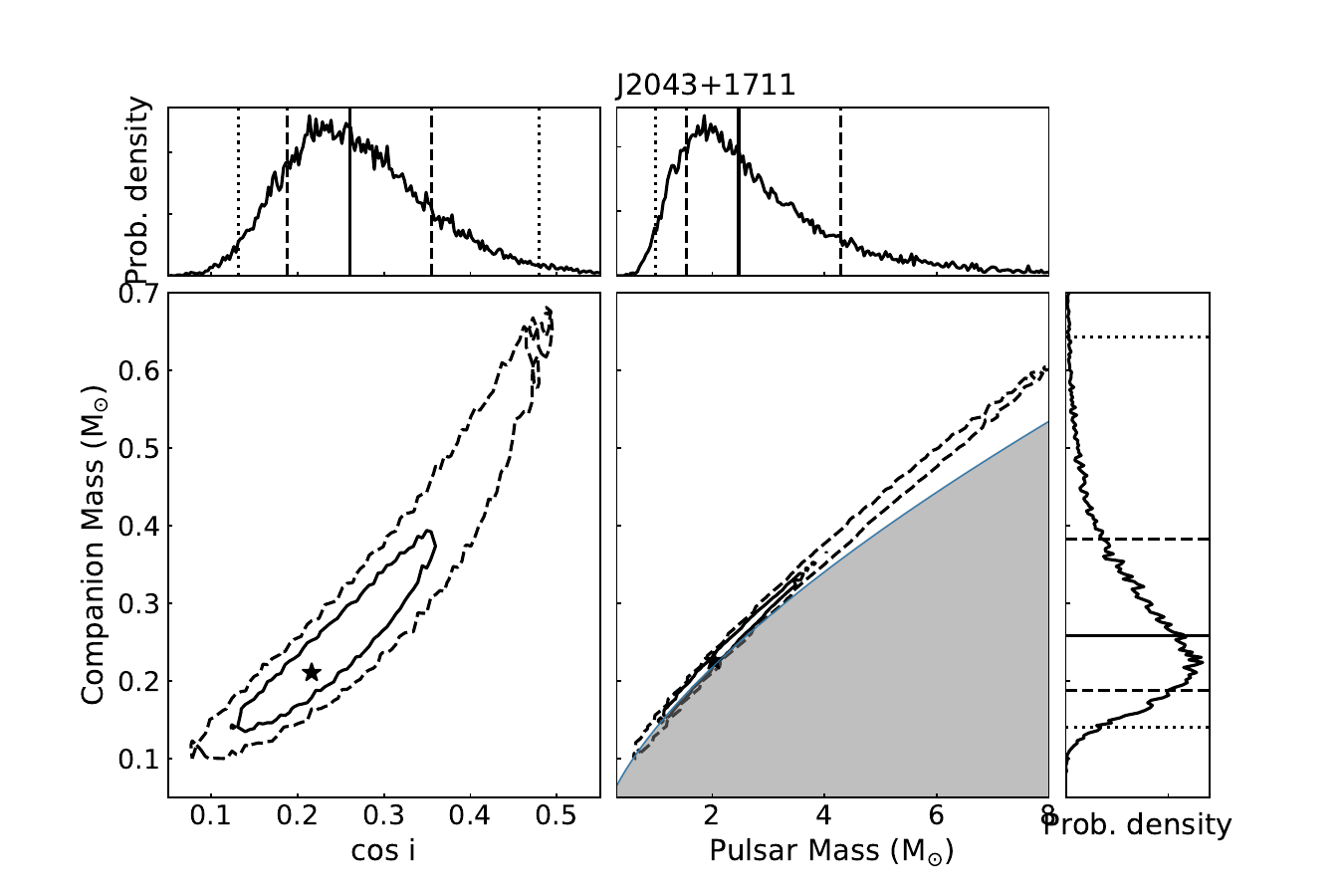}
\caption{Mass constraints from the Shapiro delay parameters $H_{\rm 3}$ and $\varsigma$ of PSRs~J0125$-$2327, J1741+1351, and J2043+1711, respectively. In the 2D posterior plots, the solid and dashed lines show 1$\sigma$ and 2$\sigma$ contours, respectively. In the side panels, the solid line represents the median value, while the dashed and dotted lines represent the 1$\sigma$ and the 2$\sigma$ bounds.
    The stars correspond to the maximum likelihood value in the 2D posterior distribution.
    \label{fig:masses}}
\end{figure}

For PSRs J0125$-$2327, J1741+1351, and J2043+1711, the timing analysis was performed using the \textsc{ELL1H} timing model, owing to their small orbital eccentricities. The Shapiro delay measurements for these three pulsars were all characterised by the $H_{\rm 3}$, $\varsigma$ parameters. From the posterior distributions of the orbital parameters, we computed the joint probability distributions of the companion mass ($m_{\rm c}$) versus the cosine of the orbital inclination angle and versus the pulsar mass ($m_{\rm p}$), respectively, using the formulas given in \citet{fw10}:
\begin{eqnarray} \label{eq:Shapiro}
    m_{\rm c} &=& \frac{H_{\rm 3}}{T_{\odot}\varsigma^3}, \\
    \left|\cos i\right| &=& \frac{1-\varsigma^2}{1+\varsigma^2}, \\
    m_{\rm p} &=& T^{1/2}_{\odot}\frac{P_{\rm b}}{2\pi}\left(\frac{m_{\rm c}\sqrt{1-\cos^2i}}{x}\right)^{3/2} - m_{\rm c}.
\end{eqnarray}

Figure~\ref{fig:masses} shows the 2D posterior distributions of the cosine of the orbital inclination angle and the pulsar mass, each plotted against the companion mass as derived from the posterior distributions of the $H_{\rm 3}$ and $\varsigma$ parameters. For all three pulsars, the companion mass constraints from the Shapiro delay are consistent with the expectations from the mass function. For J0125$-$2327, the pulsar and companion masses are constrained to be $0.762_{-0.340}^{+0.726}$\,$M_\odot$ and $0.172_{-0.065}^{+0.131}$\,$M_\odot$, respectively. While the pulsar may have a relatively low mass, the current constraints remain limited and would benefit from further improvement. For PSR J1741+1351, the pulsar and companion masses are constrained to be $1.926_{-0.611}^{+0.981}$\,$M_\odot$ and $0.319_{-0.074}^{+0.110}$\,$M_\odot$, respectively. For J2043+1711, the corresponding values are $2.472_{-0.935}^{+1.82}$\,$M_\odot$ and $0.259_{-0.071}^{+0.124}$\,$M_\odot$, which are consistent with the timing results reported in \citet{NG15+a}.

\subsection{Secular variation of orbital period and kinematic distance} \label{ssec:pbdot}

\begin{table*}
\caption{Measured orbital-period derivative alongside the expected contribution from extrinsic effects. }
\centering
\renewcommand{\arraystretch}{1.4}
\begin{tabular}{c c c c c c c c}
\hline
Pulsar name & $\dot{P}_{\rm b,Obs}$ (fs/s) & $\dot{P}_{\rm b,Shk}$ (fs/s) & $\dot{P}_{\rm b,Gal}$ (fs/s) & $\dot{P}_{\rm b,int}$ (fs/s) &$\dot{P}_{\rm b,GW}$ (fs/s) & $D_{\rm kin}$ (kpc) \\
\hline
J0125$-$2327 & $2706^{+45}_{-45}$ & $2831^{+133}_{-138}$ & $-135^{+2}_{-2}$ & $10^{+142}_{-140}$ & $-0.11^{+0.06}_{-0.19}$ & $1.24^{+0.02}_{-0.02}$ \\
J1125+7819 & $257_{-45}^{+45}$ & $2044_{-490}^{+495}$ & $-107_{-5}^{+8}$ & $631_{-658}^{+668}$ & $-0.086$ & $1.11_{-0.19}^{+0.19}$ \\
J1741+1351 & $783^{+348}_{-348}$ & $1002^{+181}_{-177}$ & $-174^{+37}_{-44}$ & $-39^{+368}_{-368}$ & $-0.097^{+0.040}_{-0.076}$ & $1.79_{-0.78}^{+0.81}$ \\
\hline
\end{tabular}
\tablefoot{The columns list the estimates for the Shklovskii effect, the contribution due to Galactic acceleration, the intrinsic orbital-period change, and the contribution from GW damping. For PSRs~J0125$-$2327 and J1741+1351, the GW damping contribution was calculated using the mass measurements obtained in Section~\ref{ssec:mass}. For PSR~J1125+7819, we assumed a pulsar mass of 1.35\,$M_\odot$ and a companion mass of 0.33\,$M_\odot$, the latter corresponding to the median mass from the mass function at an orbital inclination angle of 60\,deg. The final column lists the derived kinematic distance for each pulsar.}
\label{table:pbdot}
\end{table*}

For PSRs J0125–2327, J1125+7819, and J1741+1351, we measured the secular variation of the orbital period. In binary pulsars, the observed $\dot{P}_{\rm b}$ can arise from a combination of several effects \citep{dt91,lk05}. For the systems considered here, where tidal interactions with the white dwarf companion are expected to be negligible, the most relevant contributions can be summarised as follows:
\begin{equation} \label{eq:Pbdot_terms}
\dot{P}^{\rm obs}_{\rm b}=\dot{P}_{\rm b}^{\rm Shk}+\dot{P}_{\rm b}^{\rm Gal}+\dot{P}_{\rm b}^{\rm GR}+\dot{P}^{\dot{m}}_{\rm b}.
\end{equation}

The first term, $\dot{P}_{\rm b}^{\rm Shk}$, corresponds to the contribution caused by the transverse relative motion of the system with respect to the Solar System barycentre (SSB), commonly referred to as the Shklovskii effect. Using the measured proper motion components of the pulsar  ($\mu_{\rm\alpha}$, $\mu_{\rm\delta}$) and its distance ($d$) derived from the timing parallax, this term can be estimated via
\begin{equation} \label{eq:pbdot:Shk}
    \dot{P}^{\rm Shk}_{\rm b}=P_{\rm b}\left(\mu^2_{\rm\alpha}+\mu^2_{\rm\delta}\right)\displaystyle\frac{d}{c},
\end{equation}
where $c$ is the speed of light.

The second term, $\dot{P}_{\rm b}^{\rm Gal}$, arises from the differential acceleration due to the Galactic gravitational potential. This contribution can be computed using astrometric measurements in conjunction with a Galactic potential model \citep{dt91}. For our calculations,\footnote{The reported uncertainty here only reflects the error in the pulsar distance. A more realistic uncertainty that accounts for potential systematics in the Galactic potential model is expected to be on the order of \(10^{-15}\)–\(10^{-16}\) \citep{lgi+20}.} we adopted the timing results obtained in this work and used the numerical model described in \citet{McMillan:2017}.

The third term, \(\dot{P}_{\rm b}^{\rm GR}\), corresponds to the contribution from GW damping as predicted by General Relativity. For circular orbits, following \citet{pet64}, this term is given by\footnote{Whenever in combination with $T_\odot$, mass is understood to be in solar units.}
\begin{equation} \label{eq:pbdot:gw}
    \dot{P}_{\rm b}^{\rm GW}=-\frac{192\pi}{5}\left(\frac{2\pi}{P_{\rm b}}\right)^{5/3}\frac{(T_{\odot}m_{\rm c})^{5/3}q}{(q+1)^{1/3}},
\end{equation}
where $T_{\odot} = (\mathcal{GM})_{\odot}^\mathrm{N} / c^3$, $q\equiv m_{\rm p}/m_{\rm c}$, and $(\mathcal{GM})_{\odot}^\mathrm{N}$ denotes the solar mass parameter as defined by the IAU \citep{mamajek2015iau}.

The final term, $\dot{P}^{\dot{m}}_{\rm b}$, arises from mass loss within the system. In this work, we consider the loss of rotational energy from the pulsar to be the dominant mass-loss mechanism, as significant stellar wind from a cooled white dwarf with a strong surface gravitational field, such as the companions in the binary MSPs studied here, is not expected \citep[e.g.][]{ung08}. Following \citet{dt91}, this term can be expressed as\footnote{The uncertainty reported here includes only the measurement errors of the parameters, excluding that of $I_{\rm p}$.}
\begin{equation} \label{eq:pbdot:mdot}
    \dot{P}^{\dot{m}}_{\rm b}=\frac{8\pi^2G}{T_{\odot}c^5}\frac{I_{\rm p}}{m_{\rm p}+m_{\rm c}}\frac{\dot{P}}{P^3}P_{\rm b}.
\end{equation}
Here, $I_{\rm p}$ and $\dot{P}$ denote the moment of inertia and the intrinsic spin-period derivative of the pulsar, respectively. To estimate this contribution, we adopted a fiducial value of $I_{\rm p} = 1.4 \times 10^{45}$\,g\,cm$^2$, a typical value based on our mass measurements and different equations of state that satisfy all pulsar and LIGO/Virgo constraints \citep{ctb+20,fw24}.

Table~\ref{table:pbdot} summarises the various contributions to the observed secular variation of the orbital period. For all three pulsars, the observed $\dot{P}_{\rm b}$ is clearly dominated by the Shklovskii effect, followed by the contribution from the Galactic potential. Although the expected contribution from GW damping is well above the measurement uncertainty of the observed $\dot{P}_{\rm b}$, it remains smaller than the estimation error of the Shklovskii term, which arises primarily from the uncertainty in the parallax distance. Assuming that the observed $\dot{P}_{\rm b}$ values are entirely kinematic in origin, we derived the kinematic distances, which are also listed in Table~\ref{table:pbdot}. For PSR~J1125+7819, for which no timing- or astrometry-based distance measurement is available, we adopted a distance of 0.83\,kpc from the \textsc{YMW16} Galactic free-electron model \citep{ymw17} to compute the kinematic contributions. We note that the recently released \textsc{NE2025} model \citep{oc26} gives a very similar value of 0.87\,kpc. For this pulsar, we assumed a conservative uncertainty of 0.2\,kpc. It can be seen that for PSRs J0125$–$2327 and J1741+1351, the kinematic distances are in excellent agreement with those derived from the timing parallax (see Table~\ref{table:parallax}). Notably, the kinematic distance for PSR~J0125$–$2327 is approximately three times more precise than the parallax-based estimate. For PSR J1125+7819, the kinematic distance is slightly higher, but it remains broadly consistent with the free-electron model predictions.

\subsection{Secular variation of projected semi-major axis and annual-orbital parallax} \label{ssec:xdot}

\begin{figure*}
\centering
\includegraphics[scale=0.6]{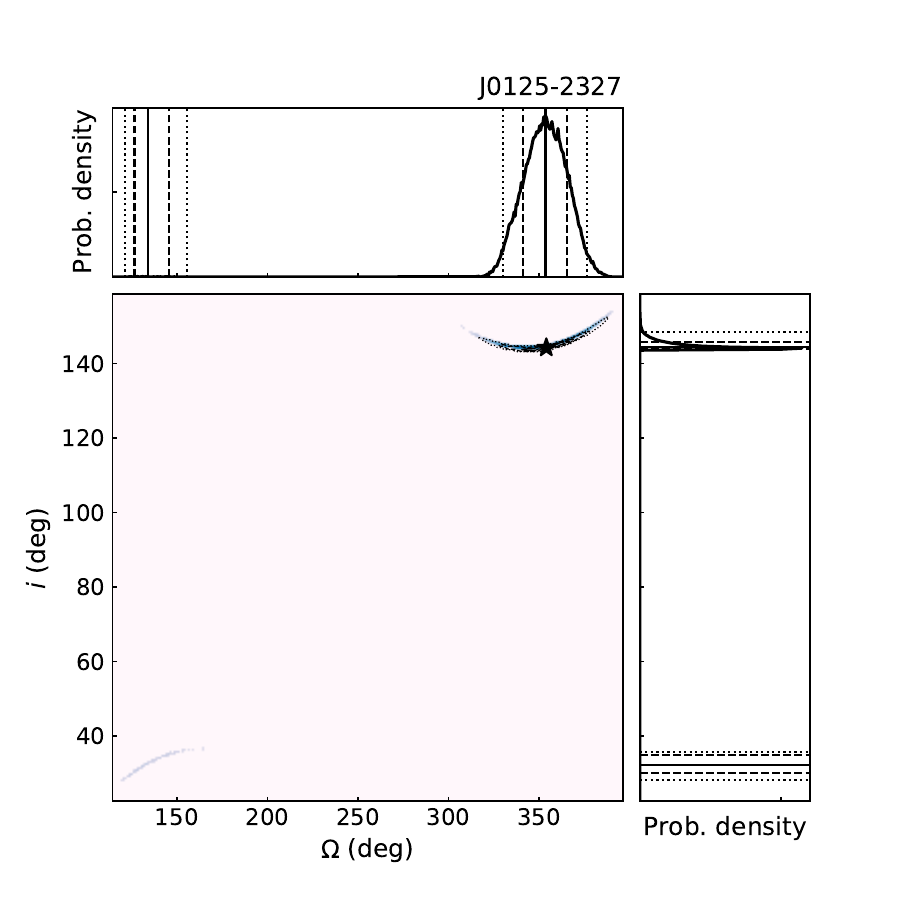}
\includegraphics[scale=0.6]{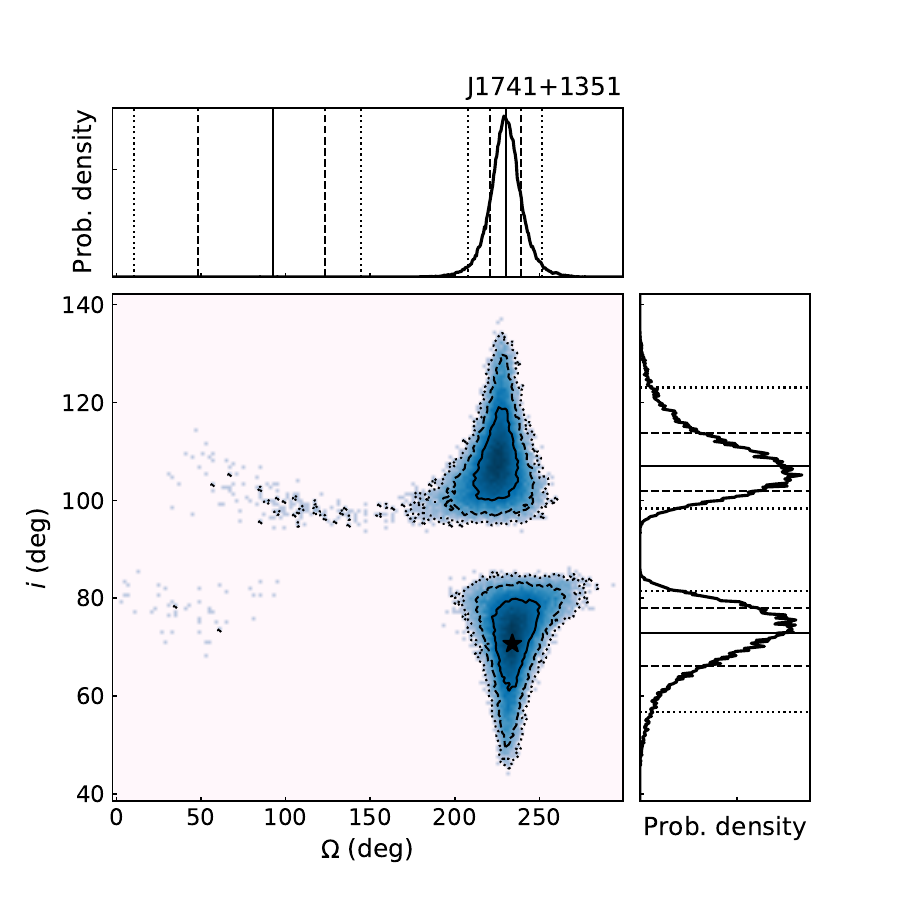}
\caption{Constraint on the longitude of ascending node and the inclination angle of the binary orbit for PSRs~J0125$-$2327 and J1741+1351, respectively, from the Bayesian sampling analysis. In the side panels, the solid line represents the median value, while the dashed and dotted lines represent the 1$\sigma$ and 2$\sigma$ confidence bounds. The maximum-likelihood values are marked with stars.\label{fig:KK}}
\end{figure*}

For PSRs J0125–2327, J1125+7819, J1741+1351, and J2019+2425, we obtained measurements of the secular variation of the projected semi-major axis, with values on the order of $10^{-14}$--$10^{-15}$. These are typical magnitudes of $\dot{x}$ caused by the proper motion of the binary pulsar system, which alters the viewing geometry of the orbit with respect to the Earth \citep[e.g.][]{zdw+19,lgi+20}. This effect can be expressed as
\citep{kop96}\footnote{Here we assumed the astronomical convention, where $\Omega$ and $i$ are different from those in \citet{kop95,kop96} as $\Omega = \pi/2 - \Omega_{\rm K96}$ and $i=\pi-i_{\rm K96}$. This applies to Eq.~\ref{eq:x_PM}--\ref{eq:x_AOP}.}
\begin{equation} \label{eq:x_PM}
    x=x_{\rm int}\left[1+\cot i(\mu_{\alpha}\cos\Omega-\mu_{\delta}\sin\Omega)(t-t_0)\right],
\end{equation}
which gives
\begin{equation} \label{eq:xdot_PM}
    \frac{\dot{x}}{x_{\rm int}}=\mu\cot i \cos (\theta_{\mu}+\Omega).
\end{equation}
Here, $\mu\equiv\sqrt{\mu^2_{\alpha}+\mu^2_{\delta}}$, $\Omega$ is the longitude of the ascending node, and $\theta_{\mu}$ is the position angle of the proper motion on the sky. In addition to this geometric effect, several other mechanisms can contribute to the observed $\dot{x}$, including the Shklovskii effect and the Galactic potential \citep{lk04}. For systems consisting of an MSP and a white dwarf companion, these contributions are typically a few orders of magnitude smaller. For instance, we estimated the Shklovskii effect and Galactic potential contribution following the same principle as for $\dot{P}_{\rm b}$. The largest Shklovskii effect contribution, in absolute value, was found in PSR J2019+2425, with a magnitude of $3\times10^{-17}$. Similarly, the largest Galactic potential contribution also occurred for PSR J2019+2425, with $\dot{x}\sim-2\times10^{-18}$. Both are well below the current measurement precision of $\dot{x}$ for these pulsars.

As an additional check, we used the formalism of \citet{dt92} to estimate the order of magnitude of the remaining terms in the $\dot x$ decomposition, namely, the aberration term due to geodetic precession ($\dot x_{\epsilon_A}$), the spin--orbit coupling of the pulsar and of the companion ($\dot x_{\rm SO}^{\rm P}$, $\dot x_{\rm SO}^{\rm C}$), and a possible third body ($\dot x_{3rd}$). The aberration contribution is suppressed by the ratio $P_{\rm p}/P_{\rm b}\sim10^{-9}$ for all four systems and is negligible ($\lesssim10^{-20}$--$10^{-22}\,\mathrm{s\,s^{-1}}$). No evidence of an additional companion was found in any of the timing solutions, so $\dot x_{3rd}$ cannot be constrained but is not expected to contribute. The spin--orbit terms, computed for the maximally unfavourable geometry ($\mathbf{s}_A\cdot\mathbf{i}=1$) and a fiducial $I_{\rm p}=1.4\times10^{45}\,\mathrm{g\,cm^2}$, are of the order of $10^{-16}$--$10^{-15}\,\mathrm{s\,s^{-1}}$, i.e. two to three orders of magnitude below the observed $\dot x\sim10^{-14}\,\mathrm{s\,s^{-1}}$, confirming that the proper-motion effect remains the dominant contribution for all four pulsars. For PSRs~J0125$-$2327 and J1741+1351, however, the (conservative) upper bound on $\dot x_{\rm SO}^{\rm P}$ reaches $\sim4\%$ of the observed $\dot x$ ($\lesssim1.6\times10^{-15}$ and $\lesssim1.0\times10^{-16}\,\mathrm{s\,s^{-1}}$, respectively), the latter being comparable to the measurement uncertainty on $\dot x$ itself. An independent constraint on the pulsar spin--orbit alignment (e.g. from polarimetric observations) would be needed to firmly rule out a non-negligible spin--orbit contribution in these two systems.

From Eq.~(\ref{eq:x_PM}), it can be seen that for binary pulsars in which the observed $\dot{x}$ is mainly attributed to the proper-motion effect, $\Omega$ can be determined with an ambiguity of $\cos(-\theta_{\mu}-\Omega)=\cos (\theta_{\mu}+\Omega)$ if the orbital inclination angle is known. In parallel, the inclination can be obtained from the Shapiro delay parameter $\sin i$, which carries a $\pi-i$ ambiguity. Consequently, this approach yields, in principle, four possible combinations of solutions for the inclination and the ascending node \citep[e.g.][]{lgi+20}. However, the annual motion of the Earth around the Sun induces an apparent periodic modulation of the orbital viewing geometry, known as the annual-orbital parallax effect \citep[e.g.][]{ehm06}. The corresponding variation in the projected semi-major axis can be expressed as \citep{kop95}
\begin{equation} \label{eq:x_AOP}
x=x_{\rm int}\left[1-\frac{\cot i}{d}(\Delta_{I_0}\cos\Omega-\Delta_{J_0}\sin\Omega)\right],
\end{equation}
where
\begin{eqnarray}
\Delta_{I_0}&=&-X\sin\alpha+Y\cos\alpha, \\
\Delta_{J_0}&=&-X\sin\delta\cos\alpha-Y\sin\delta\sin\alpha+Z\cos\delta.
\end{eqnarray}
Here, $\Vec{r}=\{X, Y, Z\}$ is the position vector of the Earth with respect to the SSB, and ($\alpha$, $\delta$) are the spherical coordinates of the barycentre of the binary system.

For well-timed binary MSPs that are relatively nearby and have sufficiently wide orbits, the annual-orbital parallax could be well measured or tightly constrained, thereby yielding a unique solution for the inclination angle and the longitude of the ascending node \citep[e.g.][]{zdw+19,rbs+24}. We investigated the annual-orbital parallax signature for PSRs~J0125$-$2327 and J1741+1351, both of which have measured $\dot{x}$ and Shapiro delay parameters. We used the \texttt{T2} binary model within \tempotwo{}, where the annual-orbital parallax effect is parameterised by the longitude of the ascending node ($\Omega$) and inclination angle ($i$). The Shapiro delay parameter, $s\equiv\sin i$, was then treated as a function of the inclination angle. To explore the $\Omega$---$i$ parameter space, we employed the \tnest{} toolkit, sampling all deterministic timing and red-noise parameters directly while fixing the white-noise parameters to their maximum-likelihood values obtained from the Bayesian timing analysis. Uniform linear priors were adopted for $\Omega$, $i$, and the companion mass $M_{\rm c}$, with prior ranges encompassing all physically allowed values.

Figure~\ref{fig:KK} shows the posterior distributions of $\Omega$ and $i$ obtained from the sampling analysis described above. For PSR~J0125$-$2327, two pairs of possible solutions were identified: $\Omega=353^{+12}_{-13}$, $i=144.4^{+1.5}_{-0.4}$ and $\Omega=134^{+12}_{-8}$, $i=32.4^{+2.5}_{-2.3}$. The logarithmic ratio of the integrated posterior mass of the former to the latter is $\simeq 2.8$, indicating that the solution with $\Omega=340^{+12}_{-13}$ and $i=144.4^{+1.5}_{-0.5}$ is strongly favoured. This provides clear evidence of the detection of the annual-orbital parallax effect in this binary MSP. For PSR~J1741+1341, the sampling converged to two regions: $\Omega=234^{+9}_{-7}$, $i=73^{+5}_{-7}$ and $\Omega=226^{+7}_{-9}$, $i=107^{+7}_{-5}$. The logarithmic ratio of the integrated posterior mass of these two solutions to the others is $\simeq 2.7$, indicating that they are strongly favoured. Meanwhile, these two regions have a nearly equal integrated posterior mass, leaving an ambiguity in the orbital inclination. Nonetheless, the two solutions yield very similar values of $\Omega$, which combined give an estimate of $\Omega=230^{+9}_{-9}$.

\subsection{Constraint on gravitational waves strain using binary orbits} \label{ssec:gw}

The precisely measured orbital period derivatives offer a valuable means to probe GWs. With orbital periods ranging from days to months, the corresponding frequency bands lie in the microhertz regime. A passing GW with a frequency matching the orbital frequency can resonantly perturb the orbit, manifesting as a secular variation in the orbital period. Assuming that all other contributions to the observed binary motion are fully accounted for, any residual variation could be attributed to such GW-induced resonance \citep{hmy13,bj22a,bj22b}. The orbital period derivatives also allowed us to probe GWs in the sub-nanohertz frequency range. By measuring the variations of the orbital period as part of the timing model fit, an upper limit can be placed on GWs in the frequency range \(1/d < f < 1/T\), where \(d\) is the observational baseline and \(T\) is the total timing span \citep{kop97}. However, in practice, it is very difficult to definitively identify changes in $\dot{P}_{\rm b}$ as an effect of interactions with GWs. At the moment, the observed $\dot{P}_{\rm b}$ can only be used to set upper limits.

\begin{figure}
\hspace*{-0.2cm}
	\includegraphics[scale=0.57]{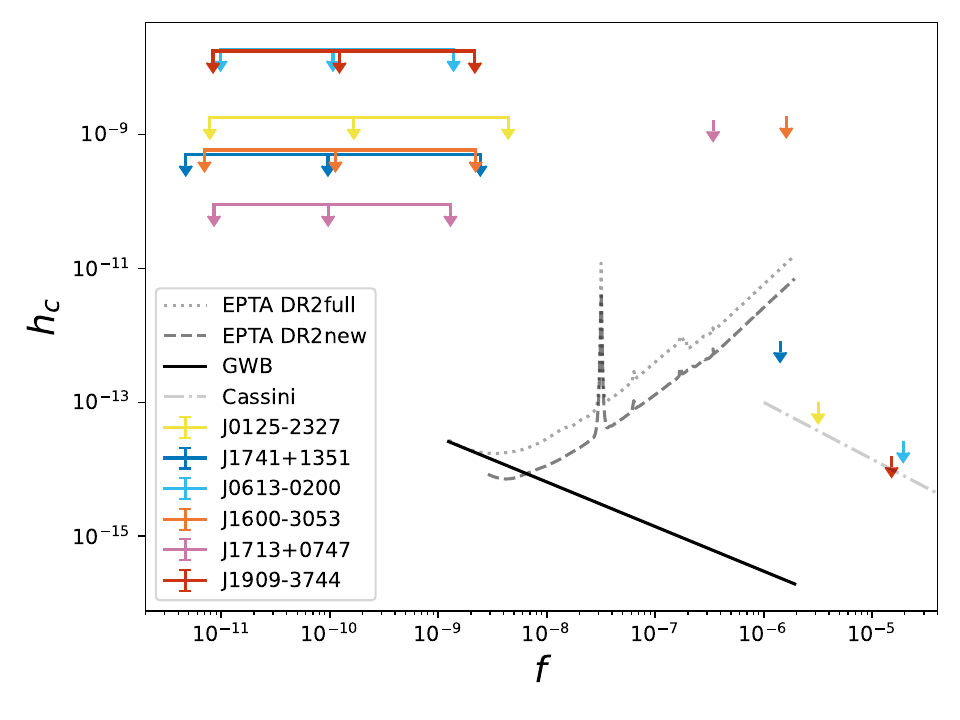}
    \caption{Upper limits on the GW strain from binary MSPs for sub-nanohertz and microhertz frequencies. The derived $\dot{P}_{\rm b,int}$ of PSR~J0125$-$2327 and J1741+1351 are from this work, while that of PSR~J1909$-$3744 is from \citet{lgi+20}, and those of the remaining pulsars are from \citet{EPTA+2023a}. For illustrative purposes, we also show the sensitivity curves obtained with the EPTA DR2 datasets \citep{EPTA+2023c} as well as the tentative GWB measurement from \citet{aaa+24}. A simplified upper limit from the Cassini spacecraft is also shown for comparison \citep{ait+03}. Note that the large upper limits for PSR~J1600$-$3053 and J1713+0747 stem from large uncertainties in the epoch of periastron or ascending node, respectively.
    \label{fig:gw_ul}}
\end{figure}

Figure~\ref{fig:gw_ul} presents the constraints on the GW background in both the sub-nanohertz and the microhertz frequency regimes, derived from binary MSPs. Here, we used the $\dot{P}_{\rm b,int}$ measurements and other quantities obtained from the dataset in this work for PSRs~J0125$-$2327 and J1741+1351 as well as those reported in \citet{lgi+20} for PSR J1909$-$3744, and in \citet{EPTA+2023a} for PSRs J0613$-$0200, J1713+0747, and J1600$-$3053. It is worth noting that for all of these pulsars, the $\dot{P}_{\rm b,int}$ values, obtained after subtracting all known contributions from the observed $\dot{P}_{\rm b,int}$, are all consistent with zero. Therefore, their measurement uncertainties were used to set the upper limits on the strength of GW. Sophisticated methods have been developed by \cite{bj22a,bj22b} to calculate the upper limits for the binary resonance frequencies. For simplicity, however, we used the results for the Hulse-Taylor binary pulsar and the scaling relation in Equation (28) from \citet{hmy13} to provide simple indications of the upper limits. In the sub-nanohertz regime the upper limits were computed at the highest frequencies corresponding to $1/d$ and set equal in the full range following Equation (71) in \cite{kop97}.

In the sub-nanohertz regime, the achievable upper limits on the GW strain are primarily determined by the orbital period. The long orbital period of PSR~J1713+0747, combined with its precise measurement of $\dot{P}_{\rm b,int}$, makes it particularly effective at constraining GWs in this frequency range. In contrast, PSRs~J0613$-$0200 and J1909$-$3744 reside in tighter binaries and thus provide significantly weaker constraints in the sub-nanohertz band. However, they yield much more stringent upper limits in the microhertz regime that are well below the sensitivity levels of current PTA experiments. The remaining pulsars, including PSRs~J0125$-$2327 and J1741+1351, deliver comparable constraints across both frequency regions.

Although a single pulsar can only constrain a very narrow range of GW frequencies, future efforts incorporating a large number of pulsars with diverse orbital periods could cover a much broader frequency spectrum. Such an application in the microhertz frequency band holds significant potential, both for detecting individual supermassive black hole binaries in their inspiralling phase and for probing cosmological sources that may emit GWs across a wide range of frequencies \citep[see, e.g., the review by][]{ska+21}.

\section{Conclusions} \label{sec:conclu}

We have performed a high-precision timing analysis of seven binary MSPs regularly monitored by the EPTA. By combining data from the five largest radio telescopes in Europe and using their coherently combined form, LEAP, we obtained phase-coherent timing solutions for all seven pulsars. The subsequent Bayesian timing and noise analysis yielded measurements of astrometric and orbital properties with broadly improved precision, some of which are reported for the first time. For PSRs~J0125$-$2327, J1741+1351, and J2043+1711, we used the astrometric measurements to derive their distances and transverse velocities, and we also obtained mass constraints through the detection of Shapiro delay signatures. Furthermore, we examined the various contributions to the secular orbital-period variations for PSRs~J0125$-$2327, J1125+7819 and J1741+1351, and we showed that under the current measurement uncertainties the observed $\dot{P}_{\rm b}$ are primarily dominated by kinematic effects. This allowed us to estimate their kinematic distances. For PSRs~J0125$-$2327 and J1741+1351, we attribute the observed secular change in the projected semi-major axis to proper-motion effects, and we identified the signature of annual-orbital parallax in their timing data. This led to a strongly favoured unique solution for the orbital ascending node ($\Omega=353^{+12}_{-13}$\,deg) and inclination angle ($i=144.4^{+1.5}_{-0.4}$\,deg) for PSRs~J0125$-$2327 and a tightly constrained ascending node $\Omega$ ($230^{+9}_{-9}$) for PSR J1741+1351. Finally, we used the orbital periods and their derivatives from several binary MSPs to place upper limits on GW signals in both the sub-nanohertz and microhertz regimes. Our results demonstrate that binary MSP orbits can provide stringent constraints in the microhertz band, complementing traditional PTA analyses.

\begin{acknowledgements}
We thank P.~Freire, T.~Tauris, L.~J.~Shao for valuable discussions on this work. This work is supported by the National Key Research and Development Programme (No. 2024YFA1611502) of China. Part of this work is based on observations with the 100-m telescope of the Max-Planck-Institut f\"ur Radioastronomie (MPIfR) at Effelsberg in Germany. The Nan\c cay Radio Observatory is operated by the Paris Observatory, associated with the French Centre National de la Recherche Scientifique (CNRS). We acknowledge financial support from the ``Action Th\'ematique de Cosmologie et Galaxies'' (ATCG), ``Action Th\'ematique Gravitation R\'ef\'erences Astronomie M\'etrologie'' (ATGRAM) and ``Action Th\'ematique Ph\'enomènes Extr\^emes et Multi-messagers'' (ATPEM) of CNRS/INSU, France. Pulsar research at the Jodrell Bank Centre for Astrophysics and the observations using the Lovell Telescope are supported by a Consolidated Grant (ST/T000414/1) from the UK's Science and Technology Facilities Council (STFC). The Sardinia Radio Telescope (SRT) is funded by the Department of University and Research (MIUR), the Italian Space Agency (ASI), and the Autonomous Region of Sardinia (RAS) and is operated as a National Facility by the National Institute for Astrophysics (INAF). The Westerbork Synthesis Radio Telescope is operated by the Netherlands Institute for Radioastronomy (ASTRON) with support from the Netherlands Foundation for Scientific Research (NWO). This work was supported by the ERC Advanced Grant ``LEAP", Grant Agreement Number 227947 (PI M.~Kramer). This work is also supported as part of the ``LEGACY'' MPG-CAS collaboration on low-frequency gravitational wave astronomy. HH acknowledges support from the PRIME programme of the German Academic Exchange Service (DAAD) with funds from the German Federal Ministry of Education and Research (BMBF). A.P acknowledges financial support from the European Research Council (ERC) starting grant 'GIGA' (grant number: 101116134) and support through the NWO-I Veni fellowship. DP acknowledges financial support from the INAF Large Grant 2023 “Gravitational Wave Detection using Pulsar Timing Arrays”.
\end{acknowledgements}

\bibliographystyle{aa}

\begin{thebibliography}{75}
\expandafter\ifx\csname natexlab\endcsname\relax\def\natexlab#1{#1}\fi

\bibitem[{{Agazie} {et~al.}(2023){Agazie}, {Alam}, {Anumarlapudi}, {Archibald}, {Arzoumanian}, {Baker}, {Blecha}, {Bonidie}, {Brazier}, {Brook}, {Burke-Spolaor}, {B{\'e}csy}, {Chapman}, {Charisi}, {Chatterjee}, {Cohen}, {Cordes}, {Cornish}, {Crawford}, {Cromartie}, {Crowter}, {Decesar}, {Demorest}, {Dolch}, {Drachler}, {Ferrara}, {Fiore}, {Fonseca}, {Freedman}, {Garver-Daniels}, {Gentile}, {Glaser}, {Good}, {G{\"u}ltekin}, {Hazboun}, {Jennings}, {Jessup}, {Johnson}, {Jones}, {Kaiser}, {Kaplan}, {Kelley}, {Kerr}, {Key}, {Kuske}, {Laal}, {Lam}, {Lamb}, {Lazio}, {Lewandowska}, {Lin}, {Liu}, {Lorimer}, {Luo}, {Lynch}, {Ma}, {Madison}, {Maraccini}, {McEwen}, {McKee}, {McLaughlin}, {McMann}, {Meyers}, {Mingarelli}, {Mitridate}, {Ng}, {Nice}, {Ocker}, {Olum}, {Panciu}, {Pennucci}, {Perera}, {Pol}, {Radovan}, {Ransom}, {Ray}, {Romano}, {Salo}, {Sardesai}, {Schmiedekamp}, {Schmiedekamp}, {Schmitz}, {Shapiro-Albert}, {Siemens}, {Simon}, {Siwek}, {Stairs}, {Stinebring}, {Stovall}, {Susobhanan}, {Swiggum}, {Taylor},
  {Turner}, {Unal}, {Vallisneri}, {Vigeland}, {Wahl}, {Wang}, {Witt}, {Young}, \& {Nanograv Collaboration}}]{NG15+a}
{Agazie}, G., {Alam}, M.~F., {Anumarlapudi}, A., {et~al.} 2023, \apjl, 951, L9

\bibitem[{{Agazie} {et~al.}(2024){Agazie}, {Antoniadis}, {Anumarlapudi}, {Archibald}, {Arumugam}, {Arumugam}, {Arzoumanian}, {Askew}, {Babak}, {Bagchi}, {Bailes}, {Bak Nielsen}, {Baker}, {Bassa}, {Bathula}, {B{\'e}csy}, {Berthereau}, {Bhat}, {Blecha}, {Bonetti}, {Bortolas}, {Brazier}, {Brook}, {Burgay}, {Burke-Spolaor}, {Burnette}, {Caballero}, {Cameron}, {Case}, {Chalumeau}, {Champion}, {Chanlaridis}, {Charisi}, {Chatterjee}, {Chatziioannou}, {Cheeseboro}, {Chen}, {Chen}, {Cognard}, {Cohen}, {Coles}, {Cordes}, {Cornish}, {Crawford}, {Cromartie}, {Crowter}, {Cury{\l}o}, {Cutler}, {Dai}, {Dandapat}, {Deb}, {DeCesar}, {DeGan}, {Demorest}, {Deng}, {Desai}, {Desvignes}, {Dey}, {Dhanda-Batra}, {Di Marco}, {Dolch}, {Drachler}, {Dwivedi}, {Ellis}, {Falxa}, {Feng}, {Ferdman}, {Ferrara}, {Fiore}, {Fonseca}, {Franchini}, {Freedman}, {Gair}, {Garver-Daniels}, {Gentile}, {Gersbach}, {Glaser}, {Good}, {Goncharov}, {Gopakumar}, {Graikou}, {Griessmeier}, {Guillemot}, {G{\"u}ltekin}, {Guo}, {Gupta}, {Grunthal}, {Hazboun},
  {Hisano}, {Hobbs}, {Hourihane}, {Hu}, {Iraci}, {Islo}, {Izquierdo-Villalba}, {Jang}, {Jawor}, {Janssen}, {Jennings}, {Jessner}, {Johnson}, {Jones}, {Joshi}, {Kaiser}, {Kaplan}, {Kapur}, {Kareem}, {Karuppusamy}, {Keane}, {Keith}, {Kelley}, {Kerr}, {Key}, {Kharbanda}, {Kikunaga}, {Klein}, {Kolhe}, {Kramer}, {Krishnakumar}, {Kulkarni}, {Laal}, {Lackeos}, {Lam}, {Lamb}, {Larsen}, {Lazio}, {Lee}, {Levin}, {Lewandowska}, {Littenberg}, {Liu}, {Liu}, {Liu}, {Lommen}, {Lorimer}, {Lower}, {Luo}, {Luo}, {Lynch}, {Lyne}, {Ma}, {Maan}, {Madison}, {Main}, {Manchester}, {Mandow}, {Mattson}, {McEwen}, {McKee}, {McLaughlin}, {McMann}, {Meyers}, {Meyers}, {Mickaliger}, {Miles}, {Mingarelli}, {Mitridate}, {Natarajan}, {Nathan}, {Ng}, {Nice}, {Ni{\c{t}}u}, {Nobleson}, {Ocker}, {Olum}, {Os{\l}owski}, {Paladi}, {Parthasarathy}, {Pennucci}, {Perera}, {Perrodin}, {Petiteau}, {Petrov}, {Pol}, {Porayko}, {Possenti}, {Prabu}, {Quelquejay Leclere}, {Radovan}, {Rana}, {Ransom}, {Ray}, {Reardon}, {Rogers}, {Romano}, {Russell},
  {Samajdar}, {Sanidas}, {Sardesai}, {Schmiedekamp}, {Schmiedekamp}, {Schmitz}, {Schult}, {Sesana}, {Shaifullah}, {Shannon}, {Shapiro-Albert}, {Siemens}, {Simon}, \& {Singha}}]{aaa+24}
{Agazie}, G., {Antoniadis}, J., {Anumarlapudi}, A., {et~al.} 2024, \apj, 966, 105

\bibitem[{Alpar {et~al.}(1982)Alpar, Cheng, Ruderman, \& Shaham}]{acrs82}
Alpar, M.~A., Cheng, A.~F., Ruderman, M.~A., \& Shaham, J. 1982, Nature, 300, 728

\bibitem[{{Antoniadis} {et~al.}(2013){Antoniadis}, {Freire}, {Wex}, {Tauris}, {Lynch}, {van Kerkwijk}, {Kramer}, {Bassa}, {Dhillon}, {Driebe}, {Hessels}, {Kaspi}, {Kondratiev}, {Langer}, {Marsh}, {McLaughlin}, {Pennucci}, {Ransom}, {Stairs}, {van Leeuwen}, {Verbiest}, \& {Whelan}}]{afw+13}
{Antoniadis}, J., {Freire}, P.~C.~C., {Wex}, N., {et~al.} 2013, Science, 340, 448

\bibitem[{Armstrong {et~al.}(2003)Armstrong, Iess, Tortora, \& Bertotti}]{ait+03}
Armstrong, J.~W., Iess, L., Tortora, P., \& Bertotti, B. 2003, ApJ, 599, 806

\bibitem[{{Babak} {et~al.}(2016){Babak}, {Petiteau}, {Sesana}, {Brem}, {Rosado}, {Taylor}, {Lassus}, {Hessels}, {Bassa}, {Burgay}, {Caballero}, {Champion}, {Cognard}, {Desvignes}, {Gair}, {Guillemot}, {Janssen}, {Karuppusamy}, {Kramer}, {Lazarus}, {Lee}, {Lentati}, {Liu}, {Mingarelli}, {Os{\l}owski}, {Perrodin}, {Possenti}, {Purver}, {Sanidas}, {Smits}, {Stappers}, {Theureau}, {Tiburzi}, {van Haasteren}, {Vecchio}, \& {Verbiest}}]{bps+16}
{Babak}, S., {Petiteau}, A., {Sesana}, A., {et~al.} 2016, \mnras, 455, 1665

\bibitem[{{Bassa} {et~al.}(2016){Bassa}, {Janssen}, {Karuppusamy}, {Kramer}, {Lee}, {Liu}, {McKee}, {Perrodin}, {Purver}, {Sanidas}, {Smits}, \& {Stappers}}]{bjk+16}
{Bassa}, C.~G., {Janssen}, G.~H., {Karuppusamy}, R., {et~al.} 2016, MNRAS, 456, 2196

\bibitem[{{Blas} \& {Jenkins}(2022{\natexlab{a}})}]{bj22b}
{Blas}, D. \& {Jenkins}, A.~C. 2022{\natexlab{a}}, \prl, 128, 101103

\bibitem[{{Blas} \& {Jenkins}(2022{\natexlab{b}})}]{bj22a}
{Blas}, D. \& {Jenkins}, A.~C. 2022{\natexlab{b}}, \prd, 105, 064021

\bibitem[{Capano {et~al.}(2020)Capano, Tews, Brown, Margalit, De, Kumar, Brown, Krishnan, \& Reddy}]{ctb+20}
Capano, C.~D., Tews, I., Brown, S.~M., {et~al.} 2020, Nature Astron., 4, 625

\bibitem[{{Cognard} {et~al.}(2013){Cognard}, {Theureau}, {Guillemot}, {Liu}, {Lassus}, \& {Desvignes}}]{ctg+13}
{Cognard}, I., {Theureau}, G., {Guillemot}, L., {et~al.} 2013, in SF2A-2013: Proceedings of the Annual meeting of the French Society of Astronomy and Astrophysics, ed. L.~{Cambresy}, F.~{Martins}, E.~{Nuss}, \& A.~{Palacios}, 327--330

\bibitem[{Damour \& Taylor(1991)}]{dt91}
Damour, T. \& Taylor, J.~H. 1991, ApJ, 366, 501

\bibitem[{Damour \& Taylor(1992)}]{dt92}
Damour, T. \& Taylor, J.~H. 1992, Phys. Rev. D, 45, 1840

\bibitem[{{Desvignes} {et~al.}(2016){Desvignes}, {Caballero}, {Lentati}, {Verbiest}, {Champion}, {Stappers}, {Janssen}, {Lazarus}, {Os{\l}owski}, {Babak}, {Bassa}, {Brem}, {Burgay}, {Cognard}, {Gair}, {Graikou}, {Guillemot}, {Hessels}, {Jessner}, {Jordan}, {Karuppusamy}, {Kramer}, {Lassus}, {Lazaridis}, {Lee}, {Liu}, {Lyne}, {McKee}, {Mingarelli}, {Perrodin}, {Petiteau}, {Possenti}, {Purver}, {Rosado}, {Sanidas}, {Sesana}, {Shaifullah}, {Smits}, {Taylor}, {Theureau}, {Tiburzi}, {van Haasteren}, \& {Vecchio}}]{dcl+16}
{Desvignes}, G., {Caballero}, R.~N., {Lentati}, L., {et~al.} 2016, MNRAS, 458, 3341

\bibitem[{Edwards {et~al.}(2006)Edwards, Hobbs, \& Manchester}]{ehm06}
Edwards, R.~T., Hobbs, G.~B., \& Manchester, R.~N. 2006, MNRAS, 372, 1549

\bibitem[{{EPTA Collaboration} {et~al.}(2023{\natexlab{a}}){EPTA Collaboration}, {Antoniadis}, {Babak}, {Bak Nielsen}, {Bassa}, {Berthereau}, {Bonetti}, {Bortolas}, {Brook}, {Burgay}, {Caballero}, {Chalumeau}, {Champion}, {Chanlaridis}, {Chen}, {Cognard}, {Desvignes}, {Falxa}, {Ferdman}, {Franchini}, {Gair}, {Goncharov}, {Graikou}, {Grie{\ss}meier}, {Guillemot}, {Guo}, {Hu}, {Iraci}, {Izquierdo-Villalba}, {Jang}, {Jawor}, {Janssen}, {Jessner}, {Karuppusamy}, {Keane}, {Keith}, {Kramer}, {Krishnakumar}, {Lackeos}, {Lee}, {Liu}, {Liu}, {Lyne}, {McKee}, {Main}, {Mickaliger}, {Ni{\c{t}}u}, {Parthasarathy}, {Perera}, {Perrodin}, {Petiteau}, {Porayko}, {Possenti}, {Quelquejay Leclere}, {Samajdar}, {Sanidas}, {Sesana}, {Shaifullah}, {Speri}, {Spiewak}, {Stappers}, {Susarla}, {Theureau}, {Tiburzi}, {van der Wateren}, {Vecchio}, {Venkatraman Krishnan}, {Verbiest}, {Wang}, {Wang}, \& {Wu}}]{EPTA+2023a}
{EPTA Collaboration}, {Antoniadis}, J., {Babak}, S., {et~al.} 2023{\natexlab{a}}, \aap, 678, A48

\bibitem[{{EPTA Collaboration} {et~al.}(2023{\natexlab{b}}){EPTA Collaboration}, {InPTA Collaboration}, {Antoniadis}, {Arumugam}, {Arumugam}, {Babak}, {Bagchi}, {Bak Nielsen}, {Bassa}, {Bathula}, {Berthereau}, {Bonetti}, {Bortolas}, {Brook}, {Burgay}, {Caballero}, {Chalumeau}, {Champion}, {Chanlaridis}, {Chen}, {Cognard}, {Dandapat}, {Deb}, {Desai}, {Desvignes}, {Dhanda-Batra}, {Dwivedi}, {Falxa}, {Ferdman}, {Franchini}, {Gair}, {Goncharov}, {Gopakumar}, {Graikou}, {Grie{\ss}meier}, {Guillemot}, {Guo}, {Gupta}, {Hisano}, {Hu}, {Iraci}, {Izquierdo-Villalba}, {Jang}, {Jawor}, {Janssen}, {Jessner}, {Joshi}, {Kareem}, {Karuppusamy}, {Keane}, {Keith}, {Kharbanda}, {Kikunaga}, {Kolhe}, {Kramer}, {Krishnakumar}, {Lackeos}, {Lee}, {Liu}, {Liu}, {Lyne}, {McKee}, {Maan}, {Main}, {Mickaliger}, {Ni{\c{t}}u}, {Nobleson}, {Paladi}, {Parthasarathy}, {Perera}, {Perrodin}, {Petiteau}, {Porayko}, {Possenti}, {Prabu}, {Quelquejay Leclere}, {Rana}, {Samajdar}, {Sanidas}, {Sesana}, {Shaifullah}, {Singha}, {Speri}, {Spiewak},
  {Srivastava}, {Stappers}, {Surnis}, {Susarla}, {Susobhanan}, {Takahashi}, {Tarafdar}, {Theureau}, {Tiburzi}, {van der Wateren}, {Vecchio}, {Venkatraman Krishnan}, {Verbiest}, {Wang}, {Wang}, \& {Wu}}]{EPTA+2023c}
{EPTA Collaboration}, {InPTA Collaboration}, {Antoniadis}, J., {et~al.} 2023{\natexlab{b}}, \aap, 678, A50

\bibitem[{{EPTA Collaboration} {et~al.}(2023{\natexlab{c}}){EPTA Collaboration}, {InPTA Collaboration}, {Antoniadis}, {Arumugam}, {Arumugam}, {Babak}, {Bagchi}, {Nielsen}, {Bassa}, {Bathula}, {Berthereau}, {Bonetti}, {Bortolas}, {Brook}, {Burgay}, {Caballero}, {Chalumeau}, {Champion}, {Chanlaridis}, {Chen}, {Cognard}, {Dandapat}, {Deb}, {Desai}, {Desvignes}, {Dhanda-Batra}, {Dwivedi}, {Falxa}, {Ferdman}, {Franchini}, {Gair}, {Goncharov}, {Gopakumar}, {Graikou}, {Grie{\ss}meier}, {Guillemot}, {Guo}, {Gupta}, {Hisano}, {Hu}, {Iraci}, {Izquierdo-Villalba}, {Jang}, {Jawor}, {Janssen}, {Jessner}, {Joshi}, {Kareem}, {Karuppusamy}, {Keane}, {Keith}, {Kharbanda}, {Kikunaga}, {Kolhe}, {Kramer}, {Krishnakumar}, {Lackeos}, {Lee}, {Liu}, {Liu}, {Lyne}, {McKee}, {Maan}, {Main}, {Mickaliger}, {Ni{\c{t}}u}, {Nobleson}, {Paladi}, {Parthasarathy}, {Perera}, {Perrodin}, {Petiteau}, {Porayko}, {Possenti}, {Prabu}, {Leclere}, {Rana}, {Samajdar}, {Sanidas}, {Sesana}, {Shaifullah}, {Singha}, {Speri}, {Spiewak}, {Srivastava},
  {Stappers}, {Surnis}, {Susarla}, {Susobhanan}, {Takahashi}, {Tarafdar}, {Theureau}, {Tiburzi}, {van der Wateren}, {Vecchio}, {Krishnan}, {Verbiest}, {Wang}, {Wang}, \& {Wu}}]{EPTA+2023b}
{EPTA Collaboration}, {InPTA Collaboration}, {Antoniadis}, J., {et~al.} 2023{\natexlab{c}}, \aap, 678, A49

\bibitem[{{Feroz} {et~al.}(2009){Feroz}, {Hobson}, \& {Bridges}}]{fhb+09}
{Feroz}, F., {Hobson}, M.~P., \& {Bridges}, M. 2009, MNRAS, 398, 1601

\bibitem[{{Freire} \& {Wex}(2010)}]{fw10}
{Freire}, P.~C.~C. \& {Wex}, N. 2010, MNRAS, 409, 199

\bibitem[{{Freire} \& {Wex}(2024)}]{fw24}
{Freire}, P. C.~C. \& {Wex}, N. 2024, Living Reviews in Relativity, 27, 5

\bibitem[{{Gravity Collaboration} {et~al.}(2021){Gravity Collaboration}, {Abuter}, {Amorim}, {Baub{\"o}ck}, {Berger}, {Bonnet}, {Brandner}, {Cl{\'e}net}, {Davies}, {de Zeeuw}, {Dexter}, {Dallilar}, {Drescher}, {Eckart}, {Eisenhauer}, {F{\"o}rster Schreiber}, {Garcia}, {Gao}, {Gendron}, {Genzel}, {Gillessen}, {Habibi}, {Haubois}, {Hei{\ss}el}, {Henning}, {Hippler}, {Horrobin}, {Jim{\'e}nez-Rosales}, {Jochum}, {Jocou}, {Kaufer}, {Kervella}, {Lacour}, {Lapeyr{\`e}re}, {Le Bouquin}, {L{\'e}na}, {Lutz}, {Nowak}, {Ott}, {Paumard}, {Perraut}, {Perrin}, {Pfuhl}, {Rabien}, {Rodr{\'\i}guez-Coira}, {Shangguan}, {Shimizu}, {Scheithauer}, {Stadler}, {Straub}, {Straubmeier}, {Sturm}, {Tacconi}, {Vincent}, {von Fellenberg}, {Waisberg}, {Widmann}, {Wieprecht}, {Wiezorrek}, {Woillez}, {Yazici}, {Young}, \& {Zins}}]{GRAVITY:2021}
{Gravity Collaboration}, {Abuter}, R., {Amorim}, A., {et~al.} 2021, Astron.~Astrophys, 647, A59

\bibitem[{{Guillemot} {et~al.}(2023){Guillemot}, {Cognard}, {van Straten}, {Theureau}, \& {G{\'e}rard}}]{gcv+23}
{Guillemot}, L., {Cognard}, I., {van Straten}, W., {Theureau}, G., \& {G{\'e}rard}, E. 2023, \aap, 678, A79

\bibitem[{{Guillemot} {et~al.}(2012){Guillemot}, {Freire}, {Cognard}, {Johnson}, {Takahashi}, {Kataoka}, {Desvignes}, {Camilo}, {Ferrara}, {Harding}, {Janssen}, {Keith}, {Kerr}, {Kramer}, {Parent}, {Ransom}, {Ray}, {Saz Parkinson}, {Smith}, {Stappers}, \& {Theureau}}]{gfc+12}
{Guillemot}, L., {Freire}, P.~C.~C., {Cognard}, I., {et~al.} 2012, MNRAS, 422, 1294

\bibitem[{{Hickish} {et~al.}(2016){Hickish}, {Abdurashidova}, {Ali}, {Buch}, {Chaudhari}, {Chen}, {Dexter}, {Domagalski}, {Ford}, {Foster}, {George}, {Greenberg}, {Greenhill}, {Isaacson}, {Jiang}, {Jones}, {Kapp}, {Kriel}, {Lacasse}, {Lutomirski}, {MacMahon}, {Manley}, {Martens}, {McCullough}, {Muley}, {New}, {Parsons}, {Price}, {Primiani}, {Ray}, {Siemion}, {van Tonder}, {Vertatschitsch}, {Wagner}, {Weintroub}, \& {Werthimer}}]{haa+16}
{Hickish}, J., {Abdurashidova}, Z., {Ali}, Z., {et~al.} 2016, Journal of Astronomical Instrumentation, 5, 1641001

\bibitem[{{Hu} {et~al.}(2020){Hu}, {Kramer}, {Wex}, {Champion}, \& {Kehl}}]{hmn+20}
{Hu}, H., {Kramer}, M., {Wex}, N., {Champion}, D.~J., \& {Kehl}, M.~S. 2020, \mnras, 497, 3118

\bibitem[{{Hui} {et~al.}(2013){Hui}, {McWilliams}, \& {Yang}}]{hmy13}
{Hui}, L., {McWilliams}, S.~T., \& {Yang}, I.-S. 2013, \prd, 87, 084009

\bibitem[{{Istrate} {et~al.}(2016){Istrate}, {Marchant}, {Tauris}, {Langer}, {Stancliffe}, \& {Grassitelli}}]{imt+16}
{Istrate}, A.~G., {Marchant}, P., {Tauris}, T.~M., {et~al.} 2016, Astron.~Astrophys, 595, A35

\bibitem[{Jacoby {et~al.}(2007)Jacoby, Bailes, Ord, Knight, \& Hotan}]{jbo+07}
Jacoby, B.~A., Bailes, M., Ord, S.~M., Knight, H.~S., \& Hotan, A.~W. 2007, ApJ, 656, 408

\bibitem[{Kopeikin(1995)}]{kop95}
Kopeikin, S.~M. 1995, ApJ, 439, L5

\bibitem[{{Kopeikin}(1996)}]{kop96}
{Kopeikin}, S.~M. 1996, ApJL, 467, L93

\bibitem[{Kopeikin(1997)}]{kop97}
Kopeikin, S.~M. 1997, Phys. Rev. D, 56, 4455

\bibitem[{{Kramer} {et~al.}(2021){Kramer}, {Stairs}, {Manchester}, {Wex}, {Deller}, {Coles}, {Ali}, {Burgay}, {Camilo}, {Cognard}, {Damour}, {Desvignes}, {Ferdman}, {Freire}, {Grondin}, {Guillemot}, {Hobbs}, {Janssen}, {Karuppusamy}, {Lorimer}, {Lyne}, {McKee}, {McLaughlin}, {M{\"u}nch}, {Perera}, {Pol}, {Possenti}, {Sarkissian}, {Stappers}, \& {Theureau}}]{ksm+21}
{Kramer}, M., {Stairs}, I.~H., {Manchester}, R.~N., {et~al.} 2021, Physical Review X, 11, 041050

\bibitem[{{Lazarus} {et~al.}(2016){Lazarus}, {Karuppusamy}, {Graikou}, {Caballero}, {Champion}, {Lee}, {Verbiest}, \& {Kramer}}]{lkg+16}
{Lazarus}, P., {Karuppusamy}, R., {Graikou}, E., {et~al.} 2016, MNRAS, 458, 868

\bibitem[{{Lentati} {et~al.}(2014){Lentati}, {Alexander}, {Hobson}, {Feroz}, {van Haasteren}, {Lee}, \& {Shannon}}]{lah+14}
{Lentati}, L., {Alexander}, P., {Hobson}, M.~P., {et~al.} 2014, MNRAS, 437, 3004

\bibitem[{{Liu} {et~al.}(2020){Liu}, {Guillemot}, {Istrate}, {Shao}, {Tauris}, {Wex}, {Antoniadis}, {Chalumeau}, {Cognard}, {Desvignes}, {Freire}, {Kehl}, \& {Theureau}}]{lgi+20}
{Liu}, K., {Guillemot}, L., {Istrate}, A.~G., {et~al.} 2020, \mnras, 499, 2276

\bibitem[{{Liu} {et~al.}(2012){Liu}, {Keane}, {Lee}, {Kramer}, {Cordes}, \& {Purver}}]{lkl+12}
{Liu}, K., {Keane}, E.~F., {Lee}, K.~J., {et~al.} 2012, MNRAS, 420, 361

\bibitem[{{Liu} {et~al.}(2011){Liu}, {Verbiest}, {Kramer}, {Stappers}, {van Straten}, \& {Cordes}}]{lvk+11}
{Liu}, K., {Verbiest}, J.~P.~W., {Kramer}, M., {et~al.} 2011, MNRAS, 417, 2916

\bibitem[{{Lorimer} \& {Kramer}(2004)}]{lk04}
{Lorimer}, D.~R. \& {Kramer}, M. 2004, {Handbook of Pulsar Astronomy}, Vol.~4 (Cambridge University Press)

\bibitem[{Lorimer \& Kramer(2005)}]{lk05}
Lorimer, D.~R. \& Kramer, M. 2005, {Handbook of Pulsar Astronomy} (Cambridge, England: Cambridge University Press)

\bibitem[{Lorimer {et~al.}(1996)Lorimer, Lyne, Bailes, Manchester, D'Amico, Stappers, Johnston, \& Camilo}]{llb+96}
Lorimer, D.~R., Lyne, A.~G., Bailes, M., {et~al.} 1996, MNRAS, 283, 1383

\bibitem[{{Lynch} {et~al.}(2018){Lynch}, {Swiggum}, {Kondratiev}, {Kaplan}, {Stovall}, {Fonseca}, {Roberts}, {Levin}, {DeCesar}, {Cui}, {Cenko}, {Gatkine}, {Archibald}, {Banaszak}, {Biwer}, {Boyles}, {Chawla}, {Dartez}, {Day}, {Ford}, {Flanigan}, {Hessels}, {Hinojosa}, {Jenet}, {Karako-Argaman}, {Kaspi}, {Leake}, {Lunsford}, {Martinez}, {Mata}, {McLaughlin}, {Noori}, {Ransom}, {Rohr}, {Siemens}, {Spiewak}, {Stairs}, {van Leeuwen}, {Walker}, \& {Wells}}]{lsk+18}
{Lynch}, R.~S., {Swiggum}, J.~K., {Kondratiev}, V.~I., {et~al.} 2018, ApJ, 859, 93

\bibitem[{Mamajek {et~al.}(2015)Mamajek, Prsa, Torres, Harmanec, Asplund, Bennett, Capitaine, Christensen-Dalsgaard, Depagne, Folkner, Haberreiter, Hekker, Hilton, Kostov, Kurtz, Laskar, Mason, Milone, Montgomery, Richards, Schou, \& Stewart}]{mamajek2015iau}
Mamajek, E.~E., Prsa, A., Torres, G., {et~al.} 2015, {IAU 2015 Resolution B3 on Recommended Nominal Conversion Constants for Selected Solar and Planetary Properties}

\bibitem[{{McEwen} {et~al.}(2020){McEwen}, {Spiewak}, {Swiggum}, {Kaplan}, {Fiore}, {Agazie}, {Blumer}, {Chawla}, {DeCesar}, {Kaspi}, {Kondratiev}, {LaRose}, {Levin}, {Lynch}, {McLaughlin}, {Mingyar}, {Noori}, {Ransom}, {Roberts}, {Schmiedekamp}, {Schmiedekamp}, {Siemens}, {Stairs}, {Stovall}, {Surnis}, \& {van Leeuwen}}]{mss+20}
{McEwen}, A.~E., {Spiewak}, R., {Swiggum}, J.~K., {et~al.} 2020, \apj, 892, 76

\bibitem[{{McMillan}(2017)}]{McMillan:2017}
{McMillan}, P.~J. 2017, MNRAS, 465, 76

\bibitem[{{Morello} {et~al.}(2022){Morello}, {Rajwade}, \& {Stappers}}]{mrs22}
{Morello}, V., {Rajwade}, K.~M., \& {Stappers}, B.~W. 2022, \mnras, 510, 1393

\bibitem[{Nice {et~al.}(2001)Nice, Splaver, \& Stairs}]{nss01}
Nice, D.~J., Splaver, E.~M., \& Stairs, I.~H. 2001, ApJ, 549, 516

\bibitem[{Nice {et~al.}(1993)Nice, Taylor, \& Fruchter}]{ntf93}
Nice, D.~J., Taylor, J.~H., \& Fruchter, A.~S. 1993, Astrophys.~J.~Let., 402, L49

\bibitem[{{Ocker} \& {Cordes}(2026)}]{oc26}
{Ocker}, S.~K. \& {Cordes}, J.~M. 2026, \apj, 1002, 3

\bibitem[{{Park} {et~al.}(2021){Park}, {Folkner}, {Williams}, \& {Boggs}}]{pfwb21}
{Park}, R.~S., {Folkner}, W.~M., {Williams}, J.~G., \& {Boggs}, D.~H. 2021, \aj, 161, 105

\bibitem[{Peters(1964)}]{pet64}
Peters, P.~C. 1964, Phys. Rev., 136, 1224

\bibitem[{Petit(2009)}]{pet09}
Petit, G. 2009, Proceedings of the International Astronomical Union, 5, 220–221

\bibitem[{Ray {et~al.}(1996)Ray, Thorsett, Jenet, van Kerkwijk, Kulkarni, Prince, Sandhu, \& Nice}]{rtj+96}
Ray, P.~S., Thorsett, S.~E., Jenet, F.~A., {et~al.} 1996, ApJ, 470, 1103

\bibitem[{{Reardon} {et~al.}(2024){Reardon}, {Bailes}, {Shannon}, {Flynn}, {Askew}, {Bhat}, {Chen}, {Cury{\l}o}, {Feng}, {Hobbs}, {Kapur}, {Kerr}, {Liu}, {Manchester}, {Mandow}, {Mishra}, {Russell}, {Shamohammadi}, {Zhang}, \& {Zic}}]{rbs+24}
{Reardon}, D.~J., {Bailes}, M., {Shannon}, R.~M., {et~al.} 2024, \apjl, 971, L18

\bibitem[{{Sesana} {et~al.}(2021){Sesana}, {Korsakova}, {Arca Sedda}, {Baibhav}, {Barausse}, {Barke}, {Berti}, {Bonetti}, {Capelo}, {Caprini}, {Garcia-Bellido}, {Haiman}, {Jani}, {Jennrich}, {Johansson}, {Khan}, {Korol}, {Lamberts}, {Lupi}, {Mangiagli}, {Mayer}, {Nardini}, {Pacucci}, {Petiteau}, {Raccanelli}, {Rajendran}, {Regan}, {Shao}, {Spallicci}, {Tamanini}, {Volonteri}, {Warburton}, {Wong}, \& {Zumalacarregui}}]{ska+21}
{Sesana}, A., {Korsakova}, N., {Arca Sedda}, M., {et~al.} 2021, Experimental Astronomy, 51, 1333

\bibitem[{{Shamohammadi} {et~al.}(2024){Shamohammadi}, {Bailes}, {Flynn}, {Reardon}, {Shannon}, {Buchner}, {Cameron}, {Camilo}, {Corongiu}, {Geyer}, {Kramer}, {Miles}, \& {Spiewak}}]{sbf+24}
{Shamohammadi}, M., {Bailes}, M., {Flynn}, C., {et~al.} 2024, \mnras, 530, 287

\bibitem[{{Shannon} {et~al.}(2025){Shannon}, {Bhat}, {Chalumeau}, {Chen}, {Cromartie}, {Gopukumar}, {Grunthal}, {Hazboun}, {Iraci}, {Joshi}, {Kato}, {Keith}, {Lee}, {Liu}, {Middleton}, {Miles}, {Mingarelli}, {Parthasarathy}, {Reardon}, {Shaifullah}, {Takahashi}, {Tiburzi}, {Truant}, {Xue}, \& {Zic}}]{sbc+25}
{Shannon}, R.~M., {Bhat}, N.~D.~R., {Chalumeau}, A., {et~al.} 2025, The Open Journal of Astrophysics, 8, 54243

\bibitem[{Shklovskii(1970)}]{shk70}
Shklovskii, I.~S. 1970, Sov. Astron., 13, 562

\bibitem[{{Smits} {et~al.}(2017){Smits}, {Bassa}, {Janssen}, {Karuppusamy}, {Kramer}, {Lee}, {Liu}, {McKee}, {Perrodin}, {Purver}, {Sanidas}, {Stappers}, \& {Zhu}}]{sbj+17}
{Smits}, R., {Bassa}, C.~G., {Janssen}, G.~H., {et~al.} 2017, Astronomy and Computing, 19, 66

\bibitem[{{Spiewak} {et~al.}(2022){Spiewak}, {Bailes}, {Miles}, {Parthasarathy}, {Reardon}, {Shamohammadi}, {Shannon}, {Bhat}, {Buchner}, {Cameron}, {Camilo}, {Geyer}, {Johnston}, {Karastergiou}, {Keith}, {Kramer}, {Serylak}, {van Straten}, {Theureau}, \& {Venkatraman Krishnan}}]{sbm+22}
{Spiewak}, R., {Bailes}, M., {Miles}, M.~T., {et~al.} 2022, \pasa, 39, e027

\bibitem[{{Stovall} {et~al.}(2019){Stovall}, {Freire}, {Antoniadis}, {Bagchi}, {Deneva}, {Garver-Daniels}, {Martinez}, {McLaughlin}, {Arzoumanian}, {Blumer}, {Brook}, {Cromartie}, {Demorest}, {DeCesar}, {Dolch}, {Ellis}, {Ferdman}, {Ferrara}, {Fonseca}, {Gentile}, {Jones}, {Lam}, {Lorimer}, {Lynch}, {Ng}, {Nice}, {Pennucci}, {Ransom}, {Spiewak}, {Stairs}, {Swiggum}, {Vigeland}, \& {Zhu}}]{sfa+19}
{Stovall}, K., {Freire}, P.~C.~C., {Antoniadis}, J., {et~al.} 2019, ApJ, 870, 74

\bibitem[{{Susarla} {et~al.}(2024){Susarla}, {Chalumeau}, {Tiburzi}, {Keane}, {Verbiest}, {Hazboun}, {Krishnakumar}, {Iraci}, {Shaifullah}, {Golden}, {Bak Nielsen}, {Donner}, {Grie{\ss}meier}, {Keith}, {Os{\l}owski}, {Porayko}, {Serylak}, {Anderson}, {Br{\"u}ggen}, {Ciardi}, {Dettmar}, {Hoeft}, {K{\"u}nsem{\"o}ller}, {Schwarz}, \& {Vocks}}]{sct+24}
{Susarla}, S.~C., {Chalumeau}, A., {Tiburzi}, C., {et~al.} 2024, \aap, 692, A18

\bibitem[{Tauris(1998)}]{tau98}
Tauris, T.~M. 1998, Astron.~Astrophys, 334, L17

\bibitem[{{Tauris} \& {van den Heuvel}(2023)}]{tv23}
{Tauris}, T.~M. \& {van den Heuvel}, E. P.~J. 2023, {Physics of Binary Star Evolution. From Stars to X-ray Binaries and Gravitational Wave Sources}

\bibitem[{Taylor(1992)}]{tay92}
Taylor, J.~H. 1992, Philos. Trans. Roy. Soc. London A, 341, 117

\bibitem[{{Unglaub}(2008)}]{ung08}
{Unglaub}, K. 2008, Astron.~Astrophys, 486, 923

\bibitem[{van Straten(2006)}]{van06}
van Straten, W. 2006, ApJ, 642, 1004

\bibitem[{van Straten {et~al.}(2012)van Straten, Demorest, \& Os{\l}owski}]{vdo12}
van Straten, W., Demorest, P., \& Os{\l}owski, S. 2012, {Astronomical Research and Technology}, 9, 237

\bibitem[{{Venkatraman Krishnan} {et~al.}(2025){Venkatraman Krishnan}, {Shao}, {Balakrishnan}, {Colom i Bernadich}, {Carleo}, {Deller}, {Corongiu}, {Freire}, {Geyer}, {Hackmann}, {Hu}, {Hu}, {Kunz}, {Kramer}, {Liu}, {Lower}, {Miao}, {Possenti}, {Perrodin}, {Pillay}, {Ransom}, {Stairs}, \& {Stappers}}]{vsb+25}
{Venkatraman Krishnan}, V., {Shao}, L., {Balakrishnan}, V., {et~al.} 2025, The Open Journal of Astrophysics, 8, 54246

\bibitem[{{Verbiest} {et~al.}(2016){Verbiest}, {Lentati}, {Hobbs}, {van Haasteren}, {Demorest}, {Janssen}, {Wang}, {Desvignes}, {Caballero}, {Keith}, {Champion}, \& et~al.}]{vlh+16}
{Verbiest}, J.~P.~W., {Lentati}, L., {Hobbs}, G., {et~al.} 2016, MNRAS, 458, 1267

\bibitem[{{Verbiest} {et~al.}(2010){Verbiest}, {Lorimer}, \& {McLaughlin}}]{vlm10}
{Verbiest}, J.~P.~W., {Lorimer}, D.~R., \& {McLaughlin}, M.~A. 2010, MNRAS, 405, 564

\bibitem[{{Verbiest} {et~al.}(2012){Verbiest}, {Weisberg}, {Chael}, {Lee}, \& {Lorimer}}]{vwc+12}
{Verbiest}, J.~P.~W., {Weisberg}, J.~M., {Chael}, A.~A., {Lee}, K.~J., \& {Lorimer}, D.~R. 2012, ApJ, 755, 39

\bibitem[{{Xu} {et~al.}(2025){Xu}, {Han}, \& {Jing}}]{xhj25}
{Xu}, J., {Han}, J., \& {Jing}, W. 2025, The Open Journal of Astrophysics, 8, 54248

\bibitem[{{Yao} {et~al.}(2017){Yao}, {Manchester}, \& {Wang}}]{ymw17}
{Yao}, J.~M., {Manchester}, R.~N., \& {Wang}, N. 2017, ApJ, 835, 29

\bibitem[{{Zhu} {et~al.}(2019){Zhu}, {Desvignes}, {Wex}, {Caballero}, {Champion}, {Demorest}, {Ellis}, {Janssen}, {Kramer}, {Krieger}, {Lentati}, {Nice}, {Ransom}, {Stairs}, {Stappers}, {Verbiest}, {Arzoumanian}, {Bassa}, {Burgay}, {Cognard}, {Crowter}, {Dolch}, {Ferdman}, {Fonseca}, {Gonzalez}, {Graikou}, {Guillemot}, {Hessels}, {Jessner}, {Jones}, {Jones}, {Jordan}, {Karuppusamy}, {Lam}, {Lazaridis}, {Lazarus}, {Lee}, {Levin}, {Liu}, {Lyne}, {McKee}, {McLaughlin}, {Os{\l}owski}, {Pennucci}, {Perrodin}, {Possenti}, {Sanidas}, {Shaifullah}, {Smits}, {Stovall}, {Swiggum}, {Theureau}, \& {Tiburzi}}]{zdw+19}
{Zhu}, W.~W., {Desvignes}, G., {Wex}, N., {et~al.} 2019, MNRAS, 482, 3249

\end{thebibliography}

\end{document}